\documentclass[twocolumn,showpacs,preprintnumbers,amsmath,amssymb]{revtex4}
\usepackage{graphicx}
\usepackage{dcolumn}
\usepackage{bm}
\makeatletter
\def\fmslash{\@ifnextchar[{\fmsl@sh}{\fmsl@sh[0mu]}}
\def\fmsl@sh[#1]#2{%
  \mathchoice
    {\@fmsl@sh\displaystyle{#1}{#2}}%
    {\@fmsl@sh\textstyle{#1}{#2}}%
    {\@fmsl@sh\scriptstyle{#1}{#2}}%
    {\@fmsl@sh\scriptscriptstyle{#1}{#2}}}
\def\@fmsl@sh#1#2#3{\m@th\ooalign{$\hfil#1\mkern#2/\hfil$\crcr$#1#3$}}
\makeatother
\newcommand{\pd}{\partial}
\newcommand{\LL}{\mathcal{L}}
\newcommand{\tr}{\mathop{\rm tr}}
\newcommand{\Tr}{\mathop{\rm Tr}}
\newcommand{\diag}{\mathop{\rm diag}}
\newcommand{\ii}{{\rm i}}
\newcommand{\Op}{\mathcal{O}}
\newcommand{\pp}{{\prime\,2}}

\newcommand{\sw}{\hat s_{0}}
\newcommand{\cw}{\hat c_{0}}
\newcommand{\vv}{\hat v_0}
\newcommand{\hc}{\text{h.c.}}
\newcommand{\eff}{\text{eff}}

\newcommand{\GeV}{{\ensuremath\rm GeV}}
\newcommand{\TeV}{{\ensuremath\rm TeV}}

\newcommand{\vA}{\mathbf{A}}
\newcommand{\vB}{\mathbf{B}}

\begin{document}
\preprint{DESY--03--167, TTP 03--24, hep-ph/0311095}

\title{The Low-Energy Structure of Little Higgs Models}

\author{Wolfgang Kilian}
 \email{wolfgang.kilian@desy.de}
\affiliation{%
Deutsches Elektronen-Synchrotron DESY,
D--22603 Hamburg, Germany
}%

\author{J\"urgen Reuter}
 \email{reuter@particle.uni-karlsruhe.de}
\affiliation{
Institut f\"ur Theoretische Teilchenphysik, University of Karlsruhe,\\
 D--76128 Karlsruhe, Germany}%

\date{\today}

\begin{abstract}
 The mechanism of electroweak symmetry breaking in Little Higgs
 Models is analyzed in an effective field theory approach. This
 enables us to identify observable effects irrespective of the
 specific structure and content of the heavy degrees of freedom.  We
 parameterize these effects in a common operator basis and present
 the complete set of anomalous contributions to gauge-boson, Higgs,
 and fermion couplings.  If the hypercharge assignments of the model
 retain their standard form, electroweak precision data are affected only
 via the $S$ and $T$ parameters and by contact interactions.  As a proof
 of principle, we apply this formalism to the minimal model and
 consider the current constraints on the parameter space.  Finally,
 we show how the interplay of measurements at LHC and a Linear
 Collider could reveal the structure of these models.
\end{abstract}


\maketitle


\section{Introduction}

In the Standard Model (SM) of electroweak interactions no symmetry
protects the Higgs boson mass from large radiative corrections.
Various scenarios have been developed which address this problem by
embedding the SM in a richer structure.  Recently, a new class of
models has been found (the Little Higgs models~\cite{LHorig,LHmin,LHmod})
where the Higgs doublet is part of a multiplet of pseudo-Goldstone
bosons.  The Goldstone-boson multiplet is associated to the
spontaneous breaking of a global symmetry at a scale $\Lambda$ which
is placed in the multi-\TeV\ range, considerably higher than the
electroweak scale~$v$.  Thus, $\Lambda$ acts as a cutoff which
separates the weakly-interacting low-energy range from a possible
strongly-interacting sector at higher energies.  The large value of
$\Lambda$ would explain the fact that no sign of such new dynamics has
yet been detected in the low-energy observables which are presently
accessible to us.

Since the Higgs bosons have interactions with gauge bosons and massive
fermions, the global symmetry can only be approximate, and the
symmetry-breaking scale $\Lambda$ cannot be arbitrarily high.
Denoting the characteristic scale of the Goldstone multiplet (the
analog of the pion decay constant) by $F$, there are
order-of-magnitude relations
\begin{equation}
  v \sim F/4\pi \sim \Lambda/16\pi^2
\end{equation}
which should be satisfied if large fine-tuning of the parameters is
excluded.

In the energy range between $F$ and $\Lambda$, Little Higgs models are
weakly-interacting models which contain, apart from the SM particles,
extra vector bosons, scalars, and fermions.  Their spectrum and
interactions are arranged in such a way that the symmetries force at
least one of the pseudo-Goldstone bosons, the Higgs particle, to be light:
$m_{H}=O(v)$.  The other new particles have masses of order $F$ (up to
several \TeV) and therefore have not yet been observed directly in
present experiments.

Nevertheless, indirect effects of the virtual exchange of heavy
particles affect the interactions of SM particles.  This fact has been
used for computing the shifts to low-energy observables and thus
to constrain Little Higgs models from precision
data~\cite{Csaki,Hewett,Han,Csaki2,CDO03}.  In the present paper, we
derive the complete low-energy effective Lagrangian, using standard
techniques of the effective-theory approach.  The Littlest Higgs model
of Ref.~\cite{LHmin} has all essential features and is thus well
suited as a concrete example, which we adopt throughout the
derivation.  For each sector of the model, we also indicate the
possible modifications that appear in the general case.

After integrating out all heavy particles, the information on the
specific model is encoded in the values of coefficients of
dimension-six operators.  This not just allows for a very simple
picture of the low-energy constraints, but gives us the opportunity to
present the complete pattern of anomalous couplings in the
gauge-boson, top-quark and Higgs sectors, which will be probed at
future colliders.  In the final section we make use of those results
and develop a strategy for reconstructing a complete model from data
at a Linear Collider in combination with LHC.

\section{Integrating Out Heavy Fields}

To get a picture of the low-energy trace of a heavy particle, one may
set up the theory in a path-integral formalism.  For instance, for two
interacting scalars $\Phi,\varphi$, where $\varphi$ is massless, the
generating functional of Green functions reads
\begin{multline}
  \mathcal{Z}[j,J]
  = \int \mathcal{D}\Phi\,\mathcal{D}\varphi
  \exp \Bigl[i\int dx\Bigl(
     \tfrac12(\pd\varphi)^2 + \tfrac12(\pd\Phi)^2 \\ - \tfrac12 M^2\Phi^2 
     - \lambda\varphi^2\Phi - \ldots + J\Phi + j\varphi\Bigr)\Bigr],
\end{multline}
where the dots indicate additional terms in the scalar potential.
The low-energy effective theory which is applicable at energy scales
$E\ll M$, is obtained by setting the source $J$ to zero (since $\Phi$
does not appear as an asymptotic state) and formally integrating out
the heavy field(s).  In the example, this is achieved by completing
the square, such that
\begin{multline}
  \frac12(\pd\Phi)^2 - \frac12 M^2\Phi^2 - \lambda\varphi^2\Phi
  = \\ -\frac12\Phi'(M^2+\pd^2)\Phi' 
  + \frac{\lambda^2}{2M^2}\varphi^2
  \left(1 + \frac{\partial^2}{M^2}\right)^{-1}\varphi^2,
\end{multline}
where
\begin{equation}
  \Phi' = \Phi + \frac{\lambda}{M^2}
   \left(1 + \frac{\partial^2}{M^2}\right)^{-1}\varphi^2,
\end{equation}
and evaluating the integral over $\Phi'$, which results in a trivial
factor.  The residual effective Lagrangian for $\varphi$ contains
virtual $\Phi$ exchange encoded as $(1+\pd^2/M^2)^{-1}$, which has to
be expanded in powers of $1/M^2$ and truncated at finite order.  

This method accounts for all tree-level effects, where terms higher
than quadratic in $\Phi$ are treated perturbatively as operator
insertions.  In particular, book-keeping is simple even if $\Phi$ gets
a vacuum expectation value or if there is nontrivial mixing between
heavy and light fields.  At loop level, there are UV-divergent
corrections to the coefficients in the effective Lagrangian which can
systematically be determined by appropriate matching conditions.  In
more general theories, there are also one-loop terms which stem from
the expansion of the Jacobi determinant, if it depends on further
light fields (e.g., gauge fields).

In the context of the minimal SM, an experimental precision at the per
mil to percent level is consistent with a truncation of the expansion
at order $1/M^2$, if $M$ is in the $\TeV$ range.  Thus, the
appropriate effective Lagrangian of Little Higgs models is given by
the SM, possibly extended by extra light Higgs multiplets, and
augmented by a small set of dimension-six
operators~\cite{BuWy86,Zep,Wudka94}.  Generically, the squared ratio
$v^2/M^2$ of the electroweak and the new-physics scales is of the same
order as $1/16\pi^2$, the prefactor of loop corrections.  In the
present context, this rough equality is dictated by naturalness.
Furthermore, while loop corrections involving only SM particles are
important, loop corrections involving heavy fields are suppressed by
additional powers of $v^2/M^2$.

The exception to this rule are loop corrections which are
quadratically divergent.  The quadratic divergence of the SM Higgs
mass is ameliorated to a logarithmic one by the matching conditions of
Little Higgs models, such that it is proportional to $M^2/(4\pi)^2\sim
v^2$.  However, for all operators of the form $(h^\dagger h)^n$ with
$n>1$, there is an uncancelled quadratic divergence which is cut off
only by the unknown UV completion of the theory.  The result is the
well-known Coleman-Weinberg potential~\cite{CW73}, which has
UV-sensitive coefficients of order one at the matching scale.

The generic suppression of radiative corrections which involve new
particles is partly reduced by logarithmic enhancement if some masses
become exceptionally light or heavy.  Such enhanced loop corrections
can have a detectable effect on observables which are very precisely
measured~\cite{Hewett,rho-loop}.  However, we should keep in mind that
there also unknown contributions from physics beyond the UV
cutoff~$\Lambda$.  These can be encoded in dimension-eight operators
and in corrections to the coefficients of dimension-six operators, and
their effect may be enhanced by the presence of new strong
interactions in the high-energy range.  Such terms are parameterically
of comparable magnitude as the loop effects due to the new heavy
particles~\cite{Wudka03}.  Thus, if we do not want to be specific
about the UV completion, we can restrict our calculation of the
low-energy effects in Little Higgs models to the coefficients of
dimension-six operators at tree-level.

\subsection{The Model}

In all Little Higgs models, the SM gauge group $SU(2)\times U(1)$ is
extended in a nontrivial way, so there are new heavy vector bosons in
the spectrum which cancel the leading cutoff dependence in the Higgs
self-energy.  After breaking of the high-energy symmetry, the heavy
states arrange themselves as massive multiplets of $SU(2)\times U(1)$.

Looking for low-energy effects, the interesting cases are triplets and
singlets of $SU(2)$ with zero hypercharge.  These vector-boson
multiplets can directly couple to the SM fermions and mix with the SM
vector bosons at leading order.  After integrating out all heavy
fields, they induce dimension-six operators in the low-energy
effective theory.  In some models~\cite{LHmod}, the extended gauge
symmetry yields additional exotic multiplets of heavy vector bosons
(e.g., $SU(2)$ doublets).  Such states may be detected by direct
observation at high-energy colliders.  However, their virtual effects
at tree level involve operators of dimension eight and higher only.
As argued before, such terms are small and compete with the low-energy
trace of the unknown UV completion, so we can consistently neglect
them.

The Littlest Higgs model~\cite{LHmin} contains exactly one extra
triplet and one extra singlet of heavy vector bosons.  The SM gauge
group is the result of the simultaneous spontaneous breaking
\begin{align}
  SU(2)_1\times SU(2)_2 &\to SU(2)
\\ 
  U(1)_1\times U(1)_2 &\to U(1).
\end{align}
This setup is easily generalized to more complicated models where
multiple vector boson triplets and singlets may exist.  In particular,
some or all extra group factors could be part of a larger simple
group.  In that case, there are relations among the gauge couplings
which restrict the allowed values of the mixing angles.  Here, we will
treat all gauge couplings as independent parameters.

We denote the $SU(2)$ and $U(1)$ gauge fields by $A_{i}^{a,\mu}$ and
$B_i^\mu$ ($i=1,2$), respectively.  In the Lagrangian, the triplet and
singlet parts are coupled by the Goldstone-boson interactions:
\begin{equation}
  \LL = \LL_0^{(3)} + \LL_0^{(1)} + \LL_0^G.
\end{equation}
The gauge-field Lagrangians are
\begin{align}
  \LL_0^{(3)} &= 
  - \sum_i\frac1{2g_i^2}\Tr\vA_{i,\mu\nu}\vA_i^{\mu\nu}
  - 2\tr A_1^\mu J^{(3)}_\mu,
  \\
  \LL_0^{(1)} &= 
  - \sum_i\frac1{2g_i^\pp}\Tr\vB_{i,\mu\nu}\vB_i^{\mu\nu}
  - \sum_i B_i^\mu J^{(1)}_{i,\mu}.
\end{align}
Here, we define the matrix-valued field strengths as
\begin{align}
  \vA^{\mu\nu}_i &= 
    \pd^\mu\vA^\nu_i - \pd^\nu\vA^\mu_i
    + i [\vA^\mu_i, \vA^\nu_i],
  \\
  \vB^{\mu\nu}_i &= 
    \pd^\mu\vB^\nu_i - \pd^\nu\vB^\mu_i,
\end{align}
with $\vA^\mu_i = A^{\mu,a}_i T^a_i$ and
$\vB^\mu_i=B^\mu_i Y_i$ ($i=1,2$).  

The vector bosons couple to the fermionic triplet and singlet currents
$J^{(3)}_\mu = J^{(3),a}_\mu\frac{\tau^a}{2}$ and $J^{(1)}_{i,\mu}$
($i=1,2$), respectively.  The triplet current is the usual left-handed
isospin current.  This current can interact with one gauge field only.  By
contrast, each singlet vector field may couple to its own fermion
current, which has the general form
\begin{equation}
\begin{split}
  J_i^{(1),\mu} &= 
    y_{L,i}\bar L_L\gamma^\mu L_L
  + y_{\nu,i}\bar\nu_R\gamma^\mu \nu_R
  + y_{\ell,i}\bar\ell_R\gamma^\mu \ell_R
\\ &
  + y_{Q,i}\bar Q_L\gamma^\mu Q_L 
  + y_{u,i}\bar u_R\gamma^\mu u_R
  + y_{d,i}\bar d_R\gamma^\mu d_R
\end{split}
\end{equation}
with \emph{a priori} arbitrary $U(1)$ charges
$y_{f,i}$~\cite{Han,Csaki2}.  The usual hypercharges are obtained
as the sum,
\begin{equation}
  y_f = \sum_i y_{f,i}
\end{equation}
for each fermion field $f$.  In order to avoid flavor-changing neutral
currents, we may assume that all $U(1)$ charges are
generation-independent.  Note that we cannot draw any conclusions from
the requirement of anomaly cancellation, since the UV completion of
the model may provide additional fermions which add to the anomalies,
but do not mix into the low-energy spectrum.

Some models~\cite{LHmod} predict $SU(2)$ triplet gauge bosons
$A^\mu_R$ which couple to triplet currents made of right-handed SM
fermions.  Since the chirality structure of the observed charged
currents is known to be left-handed to a good approximation (at least,
for the first two generations), we have to assume that these $SU(2)$
bosons are orthogonal to the left-handed $SU(2)$ sector.  Thus, the
triplet Lagrangian is unaffected up to the order we are interested in.
Nevertheless, the neutral component of a $A_R^\mu$ triplet can mix
with the hypercharge boson, so we should treat it as a $B^\mu$ boson
which generates an extra $U(1)$ symmetry.  We just have to keep in
mind that at low energies, charged $A_R^\mu$ exchange induces an extra
right-handed four-fermion contact interaction.

Turning to the Goldstone sector, let us first discuss the particular
realization of the Littlest Higgs model~\cite{LHmin}, where the gauge
group is embedded in a global $SU(5)$ group, broken down to $SO(5)$.
The representation is usually written in terms of $5\times 5$
matrices, where the $SU(2)$ generators $T^a_{1,2}$ ($a=1,2,3$) and $U(1)$
generators $Y_{1,2}$ are given by
\begin{equation}\label{generators}
  T^a_1 = \frac12
  \begin{pmatrix}
    \tau^a & & \\ & 0 & \\ & & 0
  \end{pmatrix}
  \quad\text{and}\quad
  T^a_2 = \frac12
  \begin{pmatrix}
    0 & & \\ & 0 & \\ & & -\tau^{a*}
  \end{pmatrix}
\end{equation}
and
\begin{align}
  Y_1 &= \diag(3,3,-2,-2,-2)/10, \\
  Y_2 &= \diag(2,2,2,-3,-3)/10,
\end{align}
respectively.   

The Goldstone Lagrangian describes spontaneous symmetry breaking at
the scale $F$, which is expected in the \TeV\ range.  Here, it is
parameterized by a complex symmetric $5\times 5$ matrix.  Using the
fields $H$ (light Higgs), $w^\pm,z$ (SM Goldstones), and
$\Phi^{\pm\pm},\Phi^\pm,\Phi_0,\Phi_1$ (heavy scalars) as building
blocks,
\begin{equation}
  h = 
  \begin{pmatrix}
    w^+ \\ \frac{1}{\sqrt2}(v + H + iz)
  \end{pmatrix},
  \quad
  \phi =
  \begin{pmatrix}
    \sqrt2\,\Phi^{++} & \Phi^+ \\ \Phi^+ & \Phi_0 + i\Phi_1
  \end{pmatrix}.
\end{equation}
the matrix is defined as
\begin{eqnarray}
  \Xi =& \left(\exp\frac{2i}{F}\Pi\right)\Xi_0, \quad\text{where}\quad
  \Xi_0 = 
  \begin{pmatrix}
    0 & 0 & 1 \\ 0 & 1 & 0 \\ 1 & 0 & 0
  \end{pmatrix} \nonumber \\ &
  \quad\text{and}\quad
  \Pi = \frac{1}{\sqrt2}
  \begin{pmatrix}
    0 & h & \phi \\ h^\dagger & 0 & h^T \\ \phi^\dagger & h^* & 0
  \end{pmatrix}.
\end{eqnarray}
The covariant derivative is given by
\begin{equation}
\begin{split}
  D^\mu\Xi &= \pd^\mu\Xi 
  \\ &\quad
  + i\sum_{k=1,2}\Bigl[(\vA_k^\mu\Xi + \Xi (\vA_k^\mu)^T)
  + (\vB_k^\mu\Xi + \Xi (\vB_k^\mu)^T)\Bigr],
\end{split}
\end{equation}
such that the Goldstone Lagrangian reads
\begin{equation}\label{LG}
  \LL_0^G = \frac{F^2}{8}\Tr (D_\mu\Xi)(D^\mu\Xi)^* .
\end{equation}

The generalization to more complicated models is straightforward.
There may be multiple light scalars $h$ and heavy scalars $\phi$.  In
the light sector, apart from extra Higgs doublets, there may be
singlets, triplets, etc.  The $\rho$ parameter constraints make it
unlikely that any component of a higher multiplet has a sizable vacuum
expectation value, so such extra scalars have little impact on
phenomenology~\footnote{The exception is a light singlet scalar, if it
acquires a vacuum expectation value of order $v$.  If such a scalar is
present in the spectrum, it can mix with the physical Higgs, and the
virtual effects of heavy doublets coupled to it will also
influence low-energy observables.}.  In the heavy sector, we are most
interested in triplets with hypercharge $2$ or $0$ and singlets with
hypercharge~$0$.  These can couple to Higgs doublets via
\begin{equation}
  h^\dagger \phi_2 h^\ast, \quad
  h^\dagger \phi_0 h, \quad
  h^\dagger h \sigma_0,
\end{equation}
and will induce dimension-six operators after being integrated out.
While the Littlest Higgs model contains a single $\phi_2$ multiplet,
other models realize the $\phi_0$ and $\sigma_0$ cases.

In the fermion sector of Little Higgs models, the top quark mass is
generated by mixing the known top quark $t$ with new vector-like heavy
quarks.  This interaction has the additional properties of cancelling
the quadratic cutoff dependence of the Higgs mass and generating
electroweak symmetry breaking by driving the Higgs mass squared
parameter negative.  The simplest setup involves just one heavy
vector-like fermion $T$ which is a $SU(2)$ singlet.  Many models
predict a more complicated multiplet structure.  In some cases, all
fermions have heavy partners which make them fit into multiplets of an
enlarged gauge symmetry.  However, the basic principles of
constructing the Yukawa sector~\cite{LHorig} are common to all models.

In the Littlest Higgs model, the heavy-fermion Lagrangian is built
from the chiral fields
\begin{equation}
  Q_R\!: \;
  b_R,\;
  t_R,\;
  T_R,\quad\text{and}\quad
  Q_L\!: \; 
  q_L = \begin{pmatrix} t_L \\ b_L \end{pmatrix}, \;
  T_L,
\end{equation}
namely
\begin{equation}
  \LL_t =
  \sum_{Q} \bar Q i\fmslash D Q
  + \LL_Y  - \lambda_2 F(\bar T_L T_R + \hc).
\end{equation}
The Yukawa interaction $\LL_Y$ combines $q_L$ and $T_L$ in a common
$SU(3)$ multiplet.  This implements the Little-Higgs symmetry
structure in the fermion sector, such that the leading cutoff
dependence due to top-quark loops is cancelled.  We define a $3\times
3$ matrix $\hat\chi_L^{ij} = \epsilon^{ijk}\chi_{L}^k$, where
\begin{equation}
  \chi_L = \begin{pmatrix} i\tilde q_L \\ T_L \end{pmatrix}
\quad\text{with}\quad
  \tilde q_L = i\tau^2 q_L
  = i\tau^2 \begin{pmatrix} t_L \\ b_L \end{pmatrix},
\end{equation}
and promote this to a $5\times 5$ matrix by padding zeroes:
\begin{equation}
  \hat\chi_L =
  \begin{pmatrix}
    i\tau^2 T_L & i q_L & 0 \\ -i q_L^T & 0 & 0 \\ 0 & 0 & 0
  \end{pmatrix}.
\end{equation}
With these definitions, the Yukawa interaction is given by
\begin{equation}\label{yukawa}
  \LL_Y = \lambda_1 F
    \bar t_R\Tr\left[\Xi^* (iT_2^2) \Xi^* \hat\chi_L\right] + \hc,
\end{equation}
where $T_2^2$ is the generator defined in~(\ref{generators}).

The masses of the light leptons and quarks can be generated by similar
interactions~\footnote{Flavor physics puts restrictions on the
dynamics above the scale $\Lambda$ which is responsible for generating
the fermion Yukawa interactions~\cite{Chivukula}.}, where in those
cases naturalness does not require the presence of further heavy
states if the corresponding quadratic divergences are cut off at the
scale~$\Lambda$~\cite{LHorig}.  An interesting property of the
Littlest Higgs model is the possibility to write lepton-number
violating interactions like
\begin{eqnarray}
  \LL_N =& -g_N F(\bar L^c)^{T} \Xi L,
\quad\text{where}\quad
  L = \begin{pmatrix} \tilde\ell_L \\ 0 \\ 0 \end{pmatrix}
\nonumber \\ & \quad\text{and}\quad
  \tilde\ell_L = i\tau^2\ell_L 
  = i\tau^2\ell_L \begin{pmatrix} \nu_L \\ e_L \end{pmatrix},
\end{eqnarray}
which are invariant under the full gauge symmetry.  After electroweak
symmetry breaking, such operators generate Majorana masses for
left-handed neutrinos of order $g_N v^2/F$.  Since $F$ is not large
enough to account for the huge suppression of the observed neutrino
masses, the coefficient $g_N$ must itself be small.  For instance, it
could be proportional to some power of $F/\Lambda'$, where $\Lambda'$
is a high scale where lepton number is broken.

In the general case, the construction of Yukawa interactions proceeds
along similar lines.  In at least one term, a component of the top
quark is combined with the new state(s) $T$ in a common multiplet of
the enlarged global symmetry, while there is another interaction that
generates a $T$ mass term.  This structure is consistent with the
Little-Higgs symmetry and thus allows for a sizable top-quark Yukawa
coupling without generating unwanted terms in the one-loop scalar
potential~\cite{LHorig,LHmin,LHmod}.

\subsection{Heavy Vector Fields}

Introducing the physical heavy vector bosons $X_\mu,Y_\mu$ and the SM
gauge fields $W_\mu,B_\mu$, we express the gauge fields of the
Littlest Higgs model as
\begin{align}
  A_1^\mu &= W^\mu + g_X c^2 X^\mu, &
  B_1^\mu &= B^\mu + g_Y c^\pp Y^\mu, \\
  A_2^\mu &= W^\mu - g_X s^2 X^\mu, &
  B_2^\mu &= B^\mu - g_Y s^\pp Y^\mu,
\end{align}
where
\begin{align}
  c &= \frac{g_1}{\sqrt{g_1^2 + g_2^2}}, &
  s &= \frac{g_2}{\sqrt{g_1^2 + g_2^2}},  \notag \\ 
  g_X &= \frac{g}{cs}, &
  g &= \frac{g_1g_2}{\sqrt{g_1^2 + g_2^2}}, 
\end{align}
and, analogously,
\begin{align}
  c' &= \frac{g_1'}{\sqrt{g_1^\pp + g_2^\pp}}, &
  s' &= \frac{g_2'}{\sqrt{g_1^\pp + g_2^\pp}}, \notag \\ 
  g_Y &= \frac{g'}{c's'}, &
  g' &= \frac{g_1'g_2'}{\sqrt{g_1^\pp + g_2^\pp}}, 
\end{align}
and rewrite the gauge terms in the Lagrangian:
\begin{align}
  \label{L03}
  \LL_0^{(3)} &=   
    -\frac{1}{2g^2} \tr W_{\mu\nu} W^{\mu\nu}
    - 2 \tr W^\mu J^{(3)}_\mu
    \notag \\ & \quad 
    - 2 g_X c^2 \tr X^\mu J^{(3)}_\mu
    - \frac12 \tr X_{\mu\nu} X^{\mu\nu},
  \\
  \label{L01}
  \LL_0^{(1)} &=   
    -\frac{1}{4g^\pp} B_{\mu\nu} B^{\mu\nu}
    -  B^\mu J^{(1)}_{Y,\mu}
    \notag \\ & \quad 
    - g_Y Y^\mu J^{(1)}_{\mu}
    - \frac14 Y_{\mu\nu} Y^{\mu\nu}.
\end{align}
For the matter fields $X_\mu$ and $Y_\mu$, the field strengths are
$X^{\mu\nu} = D^\mu X^\nu - D^\nu X^\mu$ (with the covariant
derivative in the adjoint representation) and $Y^{\mu\nu} = \pd^\mu
Y^\nu - \pd^\nu Y^\mu$, while the SM gauge field strengths have their
standard form, $W^{\mu\nu} = \pd^\mu W^\nu - \pd^\nu W^\mu + i[W^\mu,
W^\nu]$ and $B^{\mu\nu} = \pd^\mu B^\nu - \pd^\nu B^\mu$.

In general, the singlet currents $J^{(1)}_Y$ and $J^{(1)}$ are
linearly independent.  We express the original currents
$J^{(1)}_{1,2}$ as
\begin{subequations}
\begin{align}
  J_1^\mu &= (1-a) J_Y^\mu + J_N^\mu, \label{J1mu} \\
  J_2^\mu &= a J_Y^\mu - J_N^\mu, \label{J2mu}
\end{align}
\end{subequations}
where $J_Y$ is the canonical hypercharge current.  $J_N$ describes the
terms which deviate from the canonical hypercharge assignments.  Note that
there is some ambiguity in defining $J_N$, since we can subtract an
arbitrary multiple of $J_Y$.  This is accounted for by the parameter
$a$.  The current which is coupled to the heavy vector boson $Y$
in~(\ref{L01}) is
then given by
\begin{equation}
  J^{(1)}_{\mu} = (c^\pp - a)J^{(1)}_{Y,\mu} + J^{(1)}_{N,\mu}.
\end{equation}

Furthermore, we introduce the Higgs current
\begin{equation}\label{Vmu}
  V_\mu = i\left[h (D_\mu h)^\dagger - (D_\mu h) h^\dagger\right],
\end{equation}
which may be decomposed into its singlet and triplet parts:
\begin{equation}
  V_\mu^{(1)} = \tr{V_\mu},
\quad
  V_\mu^{(3)} = V_\mu - \frac12\tr{V_\mu}.
\end{equation}
For later use we also define field strength tensors,
\begin{equation}
  V^{(3)}_{\mu\nu} = D_\mu V^{(3)}_\nu - D_\nu V^{(3)}_\mu
\quad\text{and}\quad
  V^{(1)}_{\mu\nu} = \partial_\mu V^{(1)}_\nu - \partial_\nu V^{(1)}_\mu,
\end{equation}
where
\begin{equation}
  D_\mu V^{(3)}_\nu \equiv \partial_\mu V^{(3)}_\nu + i[W_\mu,V^{(3)}_\nu].
\end{equation}
With these definitions, the Goldstone Lagrangian (\ref{LG}) can be
expanded to yield
\begin{equation}\label{L0G}
\begin{split}
  \LL_0^G
  &= M_X^2 \tr X_\mu X^\mu
  + g_X\frac{c^2-s^2}{2}\tr [X^\mu V^{(3)}_\mu]
  \\ &\quad
  + \frac12M_Y^2 Y_\mu Y^\mu
  + g_Y\frac{c^\pp-s^\pp}{4} Y^\mu V^{(1)}_\mu
  \\ &\quad
  + \frac12\tr(D_\mu\phi)^\dagger(D^\mu\phi) 
  + (D_\mu h)^\dagger (D^\mu h)
  \\ &\quad
  - \frac{1}{6F^2} \tr \left[ V^{(3)}_\mu V^{(3),\mu} \right]
  + \ldots,
\end{split}
\end{equation}
where the omitted terms are higher-dimension interactions which are
irrelevant for our discussion.  The vector-boson masses are given by
\begin{equation}\label{XY-mass}
  M_X = g_X F / 2
\quad\text{and}\quad
  M_Y = g_Y F / (2\sqrt5) ,
\end{equation}
respectively.  After electroweak symmetry breaking, the physical
masses of the $X$ and $Y$ bosons get corrections of order $v^2/F$, but
this is irrelevant for our discussion.

Following the lines of the beginning of this section, the $X$ and $Y$
vector fields are integrated out by completing the square in the
Lagrangian.  This is achieved by the redefinitions
\begin{align}\label{square}
  X_\mu' &= X_\mu - \frac{g_X c^2}{M_X^2} J^{(3)}_\mu 
            + g_X\frac{c^2-s^2}{4M_X^2} V^{(3)}_\mu
  \\
  Y_\mu' &= Y_\mu - \frac{g_Y}{M_Y^2} J^{(1)}_{\mu}
            + g_Y\frac{c^\pp-s^\pp}{4M_Y^2} V^{(1)}_\mu
\end{align}
and leads to the low-energy effective Lagrangian
\begin{equation}\label{Leff1}
  \LL = \LL^{(3)} + \LL^{(1)},
\end{equation}
where the triplet and singlet parts are given by
\begin{align}
\begin{split}\label{L-3}
  \LL^{(3)} &= 
  - \frac{1}{2g^2} \tr [W_{\mu\nu} W^{\mu\nu}] 
  - 2\tr [W^\mu J^{(3)}_\mu]
  \\ &\quad
  + f^{(3)}_{JJ}\tr [J^{{(3)},\mu} J^{(3)}_\mu]
  + f^{(3)}_{VV} \tr [V^{(3),\mu} V_\mu]
  \\ &\quad
  + f^{(3)}_{VJ}\tr [V^{(3),\mu} J^{(3)}_\mu]
\end{split}\\
\begin{split}\label{L-1}
  \LL^{(1)} &=
  - \frac{1}{4g^\pp} B_{\mu\nu} B^{\mu\nu} 
  - B^\mu J^{(1)}_{Y,\mu}
  \\ &\quad
  + f^{(1)}_{JJ} J^{{(1)},\mu} J^{(1)}_{\mu}
  + f^{(1)}_{VV} V^{(1),\mu} V^{(1)}_\mu
  \\ &\quad
  + f^{(1)}_{VJ} V^{(1),\mu} J^{(1)}_{Y,\mu}
  + f^{(1)}_{VN} V^{(1),\mu} J^{(1)}_{N,\mu},
\end{split}
\end{align}
respectively.  The Littlest Higgs values of the coefficients are
\begin{subequations}
\begin{align}
  f^{(3)}_{JJ} &= -\frac{4c^4}{F^2},
\\ 
  f^{(3)}_{VV} &= - \frac1{6F^2} \left(1 + \frac32(c^2-s^2)^2 \right),
\\
  f^{(3)}_{VJ} &= \frac{2c^2(c^2-s^2)}{F^2},
\\
  f^{(1)}_{JJ} &= -\frac{10}{F^2},
\label{f1jjx}
\\ 
  f^{(1)}_{VV} &= -\frac5{8F^2}(c^\pp  - s^\pp)^2,
\\
  f^{(1)}_{VJ} &= \frac{5(c^\pp-a)(c^\pp-s^\pp)}{F^2}, 
\\
  f^{(1)}_{VN} &= \frac{5(c^\pp-s^\pp)}{F^2}.
\label{f1vn}
\end{align}
\end{subequations}

The overall structure of the effective Lagrangian
(\ref{Leff1}--\ref{L-1}) is generic to Little Higgs models.  In
extended models, there are extra $U(1)$ gauge symmetries which are
associated to multiple linearly independent currents $J_N$.  (In the
original version of the Littlest Higgs model~\cite{LHmin}, the extra
singlet current $J_N$ and the parameter $a$ are both zero.)  If there
are multiple Higgs doublets in the light spectrum, we can construct
multiple currents $V_{i,\mu}$.  One linear combination of these is the
Noether current of the electroweak symmetries and plays the role of
$V_\mu$ in the Littlest Higgs model, while the others provide extra
interactions which induce anomalous couplings in the multi-doublet
Higgs sector.  In the present paper, we restrict ourselves to the
discussion of a single Higgs doublet and leave the multi-doublet case
as a straightforward extension.

Otherwise, the information about the specific model is encoded in the
values of the operator coefficients.  In particular, the factor
$\sqrt5$ in the $Y$ mass~(\ref{XY-mass}) corresponds to factors of $5$
in the singlet coefficients~(\ref{f1jjx}--\ref{f1vn}).  In models with
a different $U(1)$ embedding, this prefactor will take a different
value.  The constant term in $f^{(3)}_{VV}$ is a consequence of the
nonlinear Goldstone boson representation.  The analogous constant term
in $f^{(1)}_{VV}$ happens to be zero in the Littlest Higgs model.  The
terms which involve mixing angles depend on the corresponding vector
boson spectrum.  For instance, there is a variant of the Littlest
Higgs model where the extra $U(1)$ symmetry is
ungauged~\cite{Han,Csaki2,PPP03}.  In this model, the singlet
coefficients vanish identically.

\subsection{Heavy Scalars and the Higgs Boson}

In the expansion of the Goldstone Lagrangian~(\ref{L0G}), the kinetic
energy of the heavy scalar multiplet $\phi$
\begin{equation}\label{L-phi0}
  \LL^\phi_0 = \frac12\tr(D_\mu\phi)^\dagger(D^\mu\phi)
\end{equation}
produces an extra contribution to the effective Lagrangian when $\phi$
has been integrated out.  This effective interaction has to be
combined with the other terms in the Coleman-Weinberg potential of the
scalar fields, which is generated at one-loop order.

In the Littlest Higgs model, the potential involves the Higgs doublet
$h$ and the triplet $\phi$.  The coefficients are generated by
gauge-boson and fermion exchange and are therefore proportional to the
gauge and Yukawa couplings:
\begin{equation}
\begin{split}
  \LL^{CW}_0 &= 
  - \frac12 M_\phi^2 \tr [\phi\phi^\dagger] 
  + \mu^2 (h^\dagger h) - \lambda_4 (h^\dagger h)^2  \\&\quad
  - \ii \lambda_{2\phi} 
        \left( h^\dagger \phi h^* - h^T \phi^\dagger h \right)
  - \lambda_{2\phi\phi}\tr [(\phi\phi^\dagger)(hh^\dagger)]  \\ &\quad 
  - \ii \lambda_{4\phi}(h^\dagger h)
        \left( h^\dagger \phi h^* - h^T \phi^\dagger h \right)
  - \lambda_6 (h^\dagger h)^3 
\end{split}
\end{equation}
with the $\phi$ mass parameter
\begin{equation}
   M^2_\phi = 
   -F^2 \,\left[ 
    (g_1^2 + g_2^2 + g_1^\pp + g_2^\pp) k  + \lambda_1^2 k'
   \right]
   \label{Mphi2}
\end{equation}
and the coupling constants
\begin{subequations}
\begin{align}  
  \lambda_{2\phi} &=
    \frac{F}{2 \sqrt{2}} \left[
      (g_1^2 + g_1^\pp - g_2^2 - g_2^\pp) k - \lambda_1^2 k'
    \right]
\\ 
  \lambda_4 &= M_\phi^2/4F^2
\\
  \lambda_{2\phi\phi} &= 
  - 2M_\phi^2 / 3F^2  
\\
  \lambda_{4\phi} &= 
  - {\lambda_{2\phi}}/{F^2} 
\\
  \lambda_6 &=
    - M_\phi^2/6F^4
  \label{pot-l6}
\end{align}
\end{subequations}
which are sensitive to the UV completion of the theory via the
dimensionless parameters $k$ and~$k'$.  To get the correct pattern of
electroweak symmetry breaking, the $\phi$ mass squared $M_\phi^2$ must
be positive.  This implies the relation
\begin{equation}
  \left(\frac{e^2}{s_w^2s^2c^2} + \frac{e^2}{c_w^2s^\pp c^\pp}\right)k
  + \frac{\lambda_t^2}{c_t^2}k'
  < 0
\end{equation}
which the unknown coefficients $k$ and $k'$ have to satisfy.

The Higgs mass parameter $\mu^2$ is given to leading-logarithmic
one-loop order by
\begin{align}
  \mu^2 &= 
   -\frac{3}{64 \pi^2} \left[ 
    3 g^2 M_X^2 \log \frac{\Lambda^2}{M_X^2} 
    + g^\pp M_Y^2 \log \frac{\Lambda^2}{M_Y^2}
  \right]
  \notag \\ & \quad  - \frac{\lambda}{16 \pi^2} M_\phi^2 \log
  \frac{\Lambda^2}{M_\phi^2}  
  + \frac{3\lambda_t^2}{8\pi^2} M_T^2 \log \frac{\Lambda^2}{M_T^2},
\end{align}
but there are constant one-loop and two-loop corrections to this
quantity with prefactors of the order $F^2/16\pi^2
\sim\Lambda^2/(4\pi)^4$ which are not necessarily negligible.

To get all terms that we will need later, we integrate out the heavy
scalar using the redefinition
\begin{multline}\label{phi-intout}
  \phi' = \phi 
  - \frac{2\ii\lambda_{2\phi} }{M_\phi^2}\left(
      1 + \frac{D^2}{M_\phi^2} 
      + \frac{2\lambda_{2\phi\phi}}{M_\phi^2}h h^\dagger
    \right)^{-1} \\ \times \left(
      1 + \frac{\lambda_{4\phi}}{\lambda_{2\phi}} h^\dagger h
    \right) h h^T.
\end{multline}
Expanding the resulting effective Lagrangian up to second order, we
obtain
\begin{multline}\label{L-phi}
  \LL^\phi =
  \frac{2\lambda_{2\phi}^2}{M_\phi^2}\biggl(
  (h^\dagger h)^2 
  + 2\left(
      \frac{\lambda_{4\phi}}{\lambda_{2\phi}} 
      - \frac{\lambda_{2\phi\phi}}{M_\phi^2}
    \right) (h^\dagger h)^3
  \\ + \frac{1}{M_\phi^2}\tr D_\mu(h^\ast h^\dagger) D^\mu(h
  h^T) + \ldots \biggr).
\end{multline}
The first term in this expression modifies the coefficient~$\lambda_4$,
\begin{equation}\label{l4-eff}
  \lambda_4^{\rm eff} = 
    \frac{M_\phi^2}{4F^2} - \frac{2\lambda_{2\phi}^2}{M_\phi^2}.
\end{equation}
Hence, up to corrections of order $v^4/F^2$, the Higgs mass is given by
\begin{equation}\label{Hmass}
\begin{split}
  m_H^2
    &= 2\lambda_4^{\rm eff} v^2 
\\ 
    &= -2v^2\left(\frac{e^2}{s_w^2c^2} + \frac{e^2}{c_w^2c^\pp}\right)k \,
\\ &\quad\times
      \frac{\left(\dfrac{e^2}{s_w^2s^2} + \dfrac{e^2}{c_w^2s^\pp}\right)k
            + \dfrac{\lambda_t^2}{c_t^2}k'}
           {\left(\dfrac{e^2}{s_w^2s^2c^2} 
                    + \dfrac{e^2}{c_w^2s^\pp c^\pp}\right)k
            + \dfrac{\lambda_t^2}{c_t^2}k'}.
\end{split}
\end{equation}
The $\mu$ mass parameter is related to this by $\mu^2=m_H^2/2$.
For electroweak symmetry breaking to occur, $\mu^2$ has to be
positive, so the relation
\begin{equation}\label{lambda-min}
  \frac{\lambda_{2\phi}^2}{M_\phi^4} < \frac{1}{8F^2}
\end{equation}
must be satisfied~\cite{Csaki,Han}.

The other terms in~(\ref{L-phi}) are dimension-six operators, which
may be rewritten as
\begin{multline}\label{L6phi}
  \LL_6^\phi = -
    \frac{4\lambda_{2\phi}^2}{3 F^2 M_\phi^2} (h^\dagger h)^3
    + \frac{4\lambda_{2\phi}^2}{M_\phi^4} 
    \left[
      (h^\dagger h)\left((D_\mu h)^\dagger (D^\mu h)\right) \right.
\\
    \left.
      + \left((D_\mu h)^\dagger h\right)\left(h^\dagger (D^\mu h)\right)
     \right].
\end{multline}

Again, this particular expression is specific to the Littlest Higgs
model.  However, in more general models the structure is similar:
The effective Higgs potential contains $h^4$ and $h^6$ terms,
while the exchange of heavy scalars between light Higgs bosons
generates derivative interactions.  The quantum numbers of the heavy
scalar determine the structure of these terms, i.e., the square
bracket in~(\ref{L6phi}).  Introducing the operators
\begin{align}
  \Op^{(3)}_{VV} &=  \tr{V^{(3),\mu} V^{(3)}_\mu},
  \label{opvv3} \\
  \Op_{hh} &= (h^\dagger h)\left((D_\mu h)^\dagger (D^\mu h)\right),
  \label{ophh} \\
  \Op_{h,1} &= \left((D_\mu h)^\dagger h\right)
               \left(h^\dagger(D^\mu h)\right),
  \label{oph1}  
\end{align}
from integrating out triplets with hypercharge $2$ ($\phi_2$),
hypercharge $0$ ($\phi_0$), or singlets $\sigma_0$, we obtain
interactions of the form
\begin{align}
  \phi_2 &: \Op_{hh} + \Op_{h,1},
\\
  \phi_0 &: -\Op_{VV}^{(3)} + 3\Op_{hh} - \Op_{h,1},
\\
  \sigma_0 &: -\Op_{VV}^{(3)} + 2\Op_{hh},
\end{align}
respectively.  In the Littlest Higgs model, only $\phi_2$ is present.
These derivative interactions combine with the triplet and singlet
Higgs current interactions we have encountered when integrating out
the vector fields.

\subsection{Heavy Fermions and the Top Quark}

Finally, we derive the low-energy effective Lagrangian in the
fermion sector.  We expand the Yukawa term of the Littlest Higgs model
$\LL_Y$~(\ref{yukawa}) to order $1/F^2$:
\begin{align}\label{LY2}
  \LL_Y &= 
    - \lambda_1 F\left(1 - \frac{1}{F^2}h^\dagger h\right)\bar t_R T_L
    \notag \\ & \quad + \lambda_1\sqrt2 \left(1 - \frac{2}{3F^2}h^\dagger
      h\right) \bar t_R h^T \tilde q_L
    \notag \\ & \quad - \frac{i\lambda_1}{F} \bar t_R h^\dagger\phi\tilde q_L
    + \ldots + \hc
\end{align}
Combining this with the $T$ mass term,
\begin{equation}
  \LL_T = - \lambda_2 F\bar T_R T_L + \hc,
\end{equation}
we diagonalize the two top-like states to leading order by the rotation
\begin{align}
  t_R &\to c_t t_R + s_t T_R, \\
  T_R &\to -s_t t_R + c_t T_R,
\end{align}
where the mixing angle is given by
\begin{equation}
  s_t = \frac{\lambda_1}{\sqrt{\lambda_1^2 + \lambda_2^2}},
\qquad
  c_t = \frac{\lambda_2}{\sqrt{\lambda_1^2 + \lambda_2^2}}.
\end{equation}
We may first integrate out the heavy scalar $\phi$ in the
expression~(\ref{LY2}).  Using the leading term of~(\ref{phi-intout}),
this is equivalent to the replacement
\begin{equation}
  \phi \to \frac{2\ii\lambda_{2\phi}}{M_\phi^2} h h^T.
\end{equation}
In the rotated basis, the Yukawa terms take the form
\begin{equation}
\begin{split}
  \LL_Y + \LL_T &= 
    -\frac{\lambda_t F}{c_ts_t} \bar T_R T_L
    + \lambda_t\sqrt2\,\frac{s_t}{c_t}\bar T_R h^T \tilde q_L
\\&\quad
    + \frac{\lambda_t}{F}h^\dagger h \bar t_R T_L
\\&\quad
    + \lambda_t\sqrt2\left(1 - 
        \frac{\beta}{F^2} h^\dagger h \right) \bar t_R h^T \tilde q_L
    + \hc,
\end{split}
\end{equation}
where
\begin{equation}\label{beta}
  \lambda_t = \frac{\lambda_1\lambda_2}{\sqrt{\lambda_1^2 + \lambda_2^2}}
\quad\text{and}\quad
  \beta = \frac{2}{3} - \frac{\sqrt2\lambda_{2\phi}F}{M_\phi^2}.
\end{equation}
To get the low-energy effective Lagrangian, we combine the chiral
states $T_L$ and $T_R$ to a Dirac field $T$ with mass
\begin{equation}
  M_T = \lambda_t F / c_t s_t + O(v^2/F).
\end{equation}
Completing the square in the Lagrangian,
\begin{equation}
  T' = T + \lambda_t(i\fmslash D - M_T)^{-1}
       \left(\sqrt2\,\frac{s_t}{c_t} h^T \tilde q_L
             + \frac{1}{F} h^\dagger h t_R\right),
\end{equation}
we can integrate out $T'$.  We expand the result up to the order
$1/F^2$ and obtain
\begin{align}
  \LL_f^{\rm eff}
  & = \sum_{Q=q_L,t_R,b_R} \bar Q (i\fmslash D) Q
    + \frac{2s_t^4}{F^2}\bar{\tilde q}_L h^\ast (i\fmslash D) h^T \tilde q_L
    \notag \\ & \qquad + \lambda_t\sqrt2\left(1 - 
        \frac{\beta-s_t^2}{F^2} h^\dagger h \right) 
        \left(\bar t_R h^T \tilde q_L + \hc\right).
\end{align}
Using the operator definitions
\begin{align}
  \Op_{Vq} &= \bar{\tilde q}_L\fmslash[-4mu]{V}^T \tilde q_L, \label{opvq}\\
  \Op_{Vt} &= \bar{t}_R\fmslash[-4mu]{V}^{(1)} t_R, \label{opvt}\\
  \Op_{hq} &= h^\dagger h\left(\bar t_R h^T {\tilde q}_L + \hc\right),
    \label{ophq}
\end{align}
this can be rewritten in the form
\begin{align}\label{Lf-eff}
  \LL_f^{\rm eff}
  &= \sum_{Q=b,t} \bar Q (i\fmslash D) Q
     + \lambda_t \sqrt2 \left(\bar t_R h^T {\tilde q}_L + \hc\right)
    \notag 
\\ & \qquad
 + f_{Vq}\Op_{Vq} + f_{Vt}\Op_{Vt} + f_{hq}\Op_{hq},
\end{align}
where the coefficients in the Littlest Higgs model are
\begin{equation}\label{fq}
  f_{Vq} = -\frac{s_t^4}{F^2},
\quad
  f_{Vt} = 0,
\quad
  f_{hq} = \frac{\sqrt2\lambda_t}{F^2}(c_t^2 s_t^2 - \beta),
\end{equation}
and $\beta$ is the coefficient resulting from scalar interactions
given in~(\ref{beta}).

In the effective Lagrangian~(\ref{Lf-eff}), all reference to the
specific model is encoded in the values of the coefficients $f_{Vq}$,
$f_{Vt}$, and $f_{hq}$.  Since in the Littlest Higgs model there is no
mixing of the left-handed fields, the anomalous coupling $f_{Vt}$ of
the right-handed top quark vanishes.  In general, this need not be the
case.  Furthermore, other quarks and leptons may also mix with heavy
partners.  Such mixings are constrained by the absence of
flavor-changing neutral currents.  We will not consider this
complication in the present paper.

The Lagrangian~(\ref{Lf-eff}) gives rise to the top mass
\begin{equation}
  m_t = \lambda_t v + \frac{f_{hq}}{2\sqrt2}v^3.
\end{equation}
The small correction to the canonical value $\lambda_t v$ is
detectable only if $\lambda_1$ and $\lambda_2$ are determined
directly, i.e., by measuring production and decay of the heavy $T$ at
the percent level.  This accuracy is not likely to become feasible in
the near future~\cite{PPP03}.  From the viewpoint of the low-energy
effective theory, it is more appropriate to take $m_t$ as an input
parameter and absorb the correction in the mass term.  Thus, we
rewrite (\ref{Lf-eff}) as
\begin{align}\label{Lf-effp}
  \LL_f^{\rm eff}
  &= \sum_{Q=b,t} \bar Q (i\fmslash D) Q
     + \frac{m_t}{v}\sqrt2 \left(\bar t_R h^T {\tilde q}_L + \hc\right)
    \notag 
\\ & \qquad
 + f_{Vq}\Op_{Vq} + f_{Vt}\Op_{Vt} + f_{hq}\Op_{hq}',
\end{align}
where in the redefined operator
\begin{equation}\label{ophqp}
  \Op_{hq}' =  \left(h^\dagger h - v^2/2\right)
       \left(\bar t_R h^T {\tilde q}_L + \hc\right)
\end{equation}
the contribution to the top mass is removed.

\section{Equations of Motion}

The effective Lagrangian consisting of (\ref{L-3}, \ref{L-1}) and
(\ref{L-phi}) is not yet well suited for discussing physical
observables.  The reason is the presence of couplings $V$-$J$ between
the Higgs and fermion currents, which after spontaneous symmetry
breaking induce anomalous couplings of the $W$ and $Z$ bosons to
fermions.  This is not a problem, but since gauge-boson
interactions with fermions define the gauge couplings of the SM, it is
convenient to eliminate the corrections by appropriate field
redefinitions, i.e., by applying the equations of motion.  The result
will be more transparent, and the coefficients in the effective
Lagrangian can be more easily related to measurable quantities.

It is natural to separate triplet and singlet terms in this procedure.
Here, the triplet terms conserve the approximate custodial $SU(2)_c$
symmetry of the SM, while the singlet terms (which, incidentally, are
more model-dependent) induce $SU(2)_c$ violation and thus contribute
to the $\rho$ parameter.
\begin{flushleft}
\end{flushleft}

\subsection{Custodial-\boldmath$SU(2)$ Conserving Terms}

In (\ref{L-3}), the total contribution linear in the fermionic triplet
current $J^{(3)}_\mu$ is given by
\begin{equation}\label{Jcoupling}
 \LL_J^{(3)} =
 -2\tr\left[\left(W^\mu - \frac12f_{VJ}^{(3)} V^{(3),\mu} \right)
 J^{(3)}_\mu\right].
\end{equation}
A (nonlinear) redefinition of $W_\mu$ eliminates the extra term
in~(\ref{Jcoupling}).  This is equivalent to an application of the
equations of motion, which for the $W$ field read
\begin{equation}
  0 = \frac{\delta\LL}{\delta W^\mu}
  = -\frac2{g^2} D^\nu W_{\mu\nu} + V^{(3)}_\mu - 2 J^{(3)}_\mu 
    + \ldots
\end{equation}
The omitted terms are of dimension five and higher and thus
irrelevant for our discussion.  To eliminate the $V_\mu J^\mu$ term,
we add the operator
\begin{equation}\label{opnew}
\begin{split}
  0 &= \frac12f_{VJ}^{(3)}\tr [V^{(3)}_\mu \frac{\delta\LL}{\delta W_\mu}]
\\
    &= -\frac1{g^2}f_{VJ}^{(3)}\tr [V^{(3)}_\mu D_\nu W^{\mu\nu}] 
\\ &\quad
      + \frac12 f_{VJ}^{(3)}\tr [V^{(3),\mu} V^{(3)}_\mu]
      - f_{VJ}^{(3)}\tr [V^{(3),\mu} J^{(3)}_\mu]  
\end{split}
\end{equation}
to the effective Lagrangian, such that the $V$-$J$ coupling vanishes
in the result.  Applying partial integration to the first operator on
the right-hand side of (\ref{opnew}) and combining the additional
terms with (\ref{L-3}), we obtain
\begin{align}\label{L-3new}
  \LL^{(3)} &= - \frac{1}{2g^2} \tr [W_{\mu\nu} W^{\mu\nu}] 
  - 2\tr [W^\mu J^{(3)}_\mu]
  \notag \\ & \quad 
  + f^{(3)}_{JJ} \Op^{(3)}_{JJ}
  + f_{VW} \Op_{VW}
  + f^{(3)}_{VV} \Op^{(3)}_{VV},
\end{align}
where the dimension-six operators are defined as
\begin{align}
  \Op^{(3)}_{JJ} &= \tr{J^{(3),\mu} J^{(3)}_\mu},     \label{op3jj}\\
  \Op_{VW} &= \tr{V^{(3)}_{\mu\nu} W^{\mu\nu}}, \\
  \Op^{(3)}_{VV} &=  \tr{V^{(3),\mu} V^{(3)}_\mu}.
\end{align}
The coefficients in~(\ref{L-3new}) are, in the Littlest Higgs model,
\begin{align}
  f^{(3)}_{JJ} &= -\frac{4c^4}{F^2},  \label{f3jj}\\
  f_{VW} &= -\frac{c^2(c^2-s^2)}{g^2F^2},  \label{fvw} \\
  f^{(3)}_{VV} &= \frac{(1+2c^2)(c^2-s^2)}{4F^2} - \frac{1}{6F^2}. \label{f3vv}
\end{align}

To convert this result into a more useful form, we expand the
derivative acting on $V$ and rewrite it in terms of the basis
introduced in~\cite{Zep}:
\begin{equation}\label{opVW}
  \Op_{VW} = -4\Op_W - 2\Op_{BW} - 2\Op_{WW},
\end{equation}
where
\begin{align}
  \Op_W &= i(D_\mu h)^\dagger W^{\mu\nu} (D_\nu h), \label{opw} \\
  \Op_{BW} &= -\frac12 B_{\mu\nu} h^\dagger W^{\mu\nu} h, \label{opbw}\\
  \Op_{WW} &= -\frac12 (h^\dagger h)\tr{W_{\mu\nu} W^{\mu\nu}}. \label{opww}
\end{align}
The last operator renormalizes the kinetic energy of the vector
bosons.  Noting that the gauge coupling $g$ (in our convention)
appears in the dimension-four Lagrangian only as the prefactor of the
$W$ kinetic energy, we can add a term proportional to $\tr{W_{\mu\nu}
W^{\mu\nu}}$ to the Lagrangian and completely absorb its effect into a
redefinition of $g$, a shift which is unobservable in the effective
theory.  (The same argument has been applied to the top quark mass
above.) Hence, we can replace (\ref{opVW}) by
\begin{equation}\label{opVW-expand}
  \Op_{VW} = -4\Op_W - 2\Op_{BW} - 2\Op_{WW}',
\end{equation}
where now
\begin{align}\label{opwwp}
  \Op_{WW}' &= -\frac12 (h^\dagger h - v^2/2)\tr{W_{\mu\nu} W^{\mu\nu}}.
\end{align}

The second operator on the right-hand side of~(\ref{opnew}) should
also be investigated:
\begin{equation}
  \Op_{VV}^{(3)} = \tr{V^{(3)}_\mu V^\mu}
  = \tr{V_\mu V^\mu} - \frac12\tr{V_\mu}\tr{V^\mu}.
\end{equation}
This can be rewritten as~\footnote{We choose to eliminate the operator
  $\Op_{h,2} = \frac12(\partial_\mu (h^\dagger h))^2$ of
  Ref.~\cite{Zep} in favor of $\Op_{hh}$, whose physical
  interpretation is more obvious.}
\begin{equation}
  \Op_{VV}^{(3)} = 3\Op_{hh} 
    + \frac12(h^\dagger h)((D^2h)^\dagger h + h^\dagger (D^2h)),
\end{equation}
where $\Op_{hh}$ is defined in~(\ref{ophh}).
Similar to the treatment of $\Op_{WW}$, we add a term proportional
to $(D_\mu h)^\dagger (D^\mu h)$ to the Lagrangian and absorb it in
the Higgs kinetic energy, while on the other hand we replace
$\Op_{hh}$ by
\begin{equation}\label{ophhp}
  \Op_{hh}' = (h^\dagger h-v^2/2)\left((D_\mu h)^\dagger (D^\mu h)\right).
\end{equation}
This implies a redefinition of the physical value of $v$.  From the
SM Lagrangian, the Higgs part of which reads
\begin{equation}
\begin{split}
  \LL_{\rm Higgs} &=
    (D_\mu h)^\dagger (D^\mu h) 
  + \mu^2 (h^\dagger h) - \lambda (h^\dagger h)^2
\\&\quad
  - (h^\dagger J_S + J_S^\dagger h),
\end{split}
\end{equation}
we read off the equation of motion for $h$,
\begin{equation}
  D^2 h = \mu^2 h - 2\lambda (h^\dagger h) h - J_S,
\end{equation}
where $J_S$ is the scalar current of the massive fermions which
couples to the Higgs field.  For the quartic coupling $\lambda$, we
should take the effective coupling $\lambda_4^{\rm eff}$~(\ref{l4-eff}). 
Thus, we can express $\Op_{VV}^{(3)}$ as
\begin{equation}
  \Op_{VV}^{(3)} = 
  3\Op_{hh}' - 6\lambda_4^{\rm eff} \Op_{h,3}' - \frac12\Op_{J_S}', 
\end{equation}
where the additional operators are
\begin{align}
  \Op_{h,3}' &= \frac13(h^\dagger h-v^2/2)^3, \label{oph3p}\\
  \Op_{J_S}' &= (h^\dagger h-v^2/2)(h^\dagger J_S + J_S^\dagger h).
  \label{opjsp}
\end{align}
Again, we have absorbed terms proportional to $v^2$ in the definition
of $\mu^2$, $\lambda$, and the physical fermion masses.  Additional
contributions to $\Op_{hh}'$ and $\Op_{h,3}$ come from the terms
in~(\ref{L6phi}) which encode heavy-scalar exchange.

\subsection{Custodial-\boldmath$SU(2)$ Violating Terms}

From integrating out the heavy hypercharge boson, we have obtained an
interaction of the form
\begin{equation}\label{L1-VJ}
  \LL^{(1)}_J = - \left(B^\mu - f^{(1)}_{VJ}V^{(1),\mu}\right) J^{(1)}_{Y,\mu}
  + f^{(1)}_{VN} V^{(1),\mu} J^{(1)}_{N,\mu}.
\end{equation}
Analogous to the triplet case, the coupling of $V^{(1)}$ with the
hypercharge current $J^{(1)}_Y$ can be eliminated from the effective
Lagrangian by the equations of motion.  However, if the model provides
a $U(1)$ current $J^{(1)}_N$ which is linearly independent from the
hypercharge current, the resulting extra term in~(\ref{L1-VJ}) cannot
be removed in this way.

Nevertheless, we proceed as before and add the term
\begin{align}
  0 &= f^{(1)}_{VJ} V^{(1)}_\mu \frac{\delta\LL}{\delta B_\mu}
    \notag \\ &= f^{(1)}_{VJ} V^{(1)}_\mu
      \left(-\frac{1}{g^\pp}\partial^\nu B_{\mu\nu}
            + \frac12 V_\mu^{(1)}
            - J_{Y,\mu}^{(1)}\right),
\end{align}
such that the result reads
\begin{align}
  \LL^{(1)} &= 
  - \frac{1}{4g^\pp} B_{\mu\nu} B^{\mu\nu} 
  - B^\mu J_{Y,\mu}^{(1)} 
  + f_{JJ}^{(1)}\Op^{(1)}_{JJ} 
\notag \\ & \quad
  + f_{VB}\Op_{VB} 
  + f_{VV}^{(1)}\Op_{VV}^{(1)} 
  + f^{(1)}_{VN}\Op_{VN}^{(1)}.
\end{align}
Here, the operators are defined as
\begin{align}
  \Op_{JJ}^{(1)} &= J^{(1),\mu} J^{(1)}_\mu, \label{op1jj}\\
  \Op_{VB} &= V_{\mu\nu}^{(1)} B^{\mu\nu}, \label{opvb}\\
  \Op_{VV}^{(1)} &= V^{(1),\mu} V^{(1)}_\mu, \label{op1vv} \\
  \Op_{VN}^{(1)} &= V^{(1),\mu} J^{(1)}_{N,\mu}. \label{op1vn}
\end{align}
In the Littlest Higgs model, the operator coefficients are
\begin{align}
  f_{JJ}^{(1)} &= -\frac{10}{F^2}, \label{f1jj} \\
  f_{VB} &= -\frac{5(c^\pp-a)(c^\pp-s^\pp)}{2g^\pp F^2}, \label{f1vb} \\
  f_{VV}^{(1)} &= \frac{5(1+2c^\pp-4a)(c^\pp-s^\pp)}{8F^2},
    \label{f1vv} \\
  f^{(1)}_{VN} &= \frac{5(c^\pp-s^\pp)}{F^2}.
\end{align}
Switching to a more familiar basis, we expand the operators as follows:
\begin{align}
  \Op_{VB} &= -8\Op_{B} - 4\Op_{BW} - 4\Op_{BB}', \\
  \Op_{VV}^{(1)} &= 2\Op_{hh}' -12\lambda_4^{\rm eff}\Op_{h,3}' 
     - \Op_{J_S}' + 4\Op_{h,1}
\end{align}
where the new terms are
\begin{align}
  \Op_B  &= \frac{i}{2} (D_\mu h)^\dagger (D_\nu h) B^{\mu\nu},
  \label{opb}
\\
  \Op_{BB}' &= -\frac14 (h^\dagger h - v^2/2) B_{\mu\nu} B^{\mu\nu},
    \label{opbbp}
\\
  \Op_{h,1}' &= \left((D_\mu h)^\dagger h\right)
               \left(h^\dagger(D^\mu h)\right)
              - (v^2/2) (D_\mu h^\dagger)(D^\mu h). \label{oph1p}
\end{align}
Analogous to $\Op_{hh}'$, in the definition of the operator $\Op_{h,1}'$
the contribution which would modify the Higgs kinetic energy in
unitary gauge has been subtracted and absorbed in the
definition of~$v$.  Finally, we note that from~(\ref{L6phi}) we get an
additional contribution to the coefficient of the $SU(2)_c$-violating
operator $\Op_{h,1}$.

\section{Precision Observables}

In the previous sections, we have derived the low-energy effective
Lagrangian of a Little Higgs model, which is applicable in the energy
range below the lowest-lying new particle in the spectrum.  Collecting
all terms, we list the complete result in the appendix as
Eq.~(\ref{L6}).  

Below we will discuss the contributions to the electroweak precision
observables which follow from this expression.  Anomalous vector-boson
and Higgs couplings are the subject of the next section.

\subsection{Oblique Corrections}

The two operators $\Op_{BW}$ and $\Op_{h,1}$ influence the gauge-boson
two-point functions.  These corrections are usually expressed in terms
of the $S,T,U$ parameters~\cite{STU,Georgi91,Zep}.  In our context,
there is no dimension-six operator which corresponds to $U$, so
$\Delta U$ is zero.  The other two parameters get contributions from
the exchange of heavy particles:

Expanding the operator
\begin{equation}
  \Op_{BW} = -\frac12 B_{\mu\nu} h^\dagger W^{\mu\nu} h
\end{equation}
in terms of physical fields, we have to modify the rotation of neutral
fields by the correction present in~(\ref{L6})
\begin{align}
  W^3 &= \frac{e}{s_w}\left[c_w \left(1 + M_Z^2 (f_{VW}+2f_{VB})\right) Z 
               + s_w A\right],
\\
  B  &= \frac{e}{c_w}\left[-s_w \left(1 + M_Z^2 (f_{VW}+2f_{VB})\right) Z 
                + c_w A\right],
\end{align}
in order to get the correct kinetic energies of $Z$ and $A$ in the effective
Lagrangian.  Correspondingly, the gauge couplings $g$ and $g'$ are
given by
\begin{align}
  g &= \frac{e}{s_w}(1 + M_W^2 (f_{VW}+2f_{VB})), \label{geff}
\\
  g' &= \frac{e}{c_w}\left(1 + [M_Z^2-M_W^2](f_{VW}+2f_{VB})\right),
\end{align}
if expressed in terms of $e$ and $s_w,c_w$.

Here, $e$ is the ordinary electromagnetic coupling.  (In practice, we
have to account for a nontrivial scale dependence in this quantity,
but this effect is universal and independent of our discussion.)  For
the definition of the weak mixing angle $s_w$, we first consider the
special case where $J_N=0$, i.e., the hypercharge vector bosons couple
only to the standard hypercharge current.  This covers, in particular,
the original Littlest Higgs model where the fermions are gauged only
under one $U(1)$ group.  Then, the sine of the weak mixing angle $s_w$
is measured directly in $Z$ decays, since in our framework both the
vector and the axial vector coupling receive the same correction,
\begin{equation}\label{gzeff}
  \Delta v_f/v_f = \Delta a_f/a_f = M_Z^2 (f_{VW} + 2f_{VB}),
\end{equation}
such that the ratio $v_f/a_f$ is unaffected.

As a result, $\Op_{BW}$ contributes to the $S$ parameter. In our case,
we have
\begin{align}
\label{deltaS}
  \Delta S &= 8\pi v^2 (f_{VW} + 2f_{VB})
  \\ &
  = - 8\pi c^2(c^2-s^2)\frac{v^2}{g^2 F^2}
  \notag\\ &\quad
    -40\pi (c^\pp-a)(c^\pp-s^\pp)\frac{v^2}{g^\pp F^2}.
\end{align}
The second equation gives the value in the Littlest Higgs model.

Turning to the $SU(2)_c$-violating sector, the operator
$\Op_{h,1}'$~(\ref{oph1p}) 
yields a correction to the $W$ mass (but not the $Z$ mass): 
\begin{equation}
  \Delta M_W^2/M_W^2 = -\frac{v^2}{2}f_{h,1}.
\end{equation}
This is equivalent to a contribution to the $T$~parameter:
\begin{align}\label{deltaT}
  \alpha\Delta T &= -\frac{v^2}{2}f_{h,1}
\\ &
  = -\frac54(1+2c^\pp-4a)(c^\pp-s^\pp)\frac{v^2}{F^2}
    -\frac{2v^2\lambda_{2\phi}^2}{M_\phi^4},
\end{align}
where again the second line applies to the Littlest Higgs model only.

Collecting all contributions, the physical vector masses get shifted
as follows:
\begin{equation}\label{mw-mz}
  M_W^2 = \left(\frac{ev}{2s_w}\right)^2 (1 + x),
\qquad
  M_Z^2 = \left(\frac{ev}{2s_w c_w}\right)^2 (1 + y),
\end{equation}
where
\begin{align}\label{x}
  x &= \alpha\left(\frac{\Delta S}{4s_w^2} + \Delta T\right)
  = 2M_W^2 (f_{VW} + 2f_{VB}) 
    -\frac{v^2}{2}f_{h,1},
\\
  \label{y}
  y &= \alpha\left(\frac{\Delta S}{4s_w^2c_w^2}\right)
  = 2M_Z^2 (f_{VW} + 2f_{VB}).
\end{align}

\subsection{Non-Universal Hypercharges}

If the model contains a current $J_N$ which is linearly independent
from the hypercharge current $J_Y$, the situation becomes more
complicated.  This typically happens if the fermions are charged under
more than one $U(1)$ gauge group, since there is no particular reason
to have the two $U(1)$ charges proportional to each other.

We may use the freedom of choosing the parameter $a$ in (\ref{J1mu},
\ref{J2mu}) to remove, for instance, the left-handed lepton
contribution in $J_N$.  Then, the unitary-gauge interactions
induced by $\Op_{VN}$ are
\begin{equation}
\begin{split}
  \LL_{VN} &=
  \frac{-2M_W^2}{gc_w}f_{VN}\left(
    z_\ell\bar\ell_R\fmslash Z\ell_R
   + z_Q\bar Q_L\fmslash Z Q_L \right.
\\ &\quad \left. \qquad\qquad
   + z_u\bar u_R\fmslash Z u_R
   + z_d\bar u_R\fmslash Z d_R
  \right) + \ldots,
\end{split}
\end{equation}
with some fixed parameters $z_f$, where the dots indicate couplings
which involve the Higgs field.  To verify that the terms in $\LL_{VN}$
cannot be eliminated, we recall that the couplings to both the neutral
isospin and hypercharge currents satisfy the sum rules
\begin{align}
  g_L^\nu + g_L^\ell &= g_R^\nu + g_R^\ell,
\\
  g_L^u + g_L^d &= g_R^u + g_R^d,
\\
  g_L^\nu + g_L^\ell &= -3(g_L^u + g_L^d).
\end{align}
Any linear combination of the two currents also satisfies these sum
rules.  In particular, this holds for the electromagnetic current and
for the current coupled to the $Z$-boson.  Higher-dimensional bosonic
operators do not affect this property.

However, by definition, the sum rules are violated by a nonvanishing
$J_N$.  Therefore, its presence can be constrained, e.g., by measuring
the ratios
\begin{equation}
  r_\ell = \frac{g_R^\nu + g_R^\ell}{g_L^\nu + g_L^\ell},
\qquad
  r_{q} = \frac{g_R^u + g_R^d}{g_L^u + g_L^d},
\qquad
  r_{q\ell} = -3\frac{g_L^u + g_L^d}{g_L^\nu + g_L^\ell}.
\end{equation}
In other words, if any of these quantities deviates from unity, we
know that the model contains a third linearly independent current,
which in the present context is due to non-universal charge
assignments for the $U(1)$ gauge groups.

In this situation, the standard two-parameter analysis of the
electroweak precision observables is no longer appropriate.  If the
extra $U(1)$ charges (i.e., the parameters $z_\ell,z_Q,z_u,z_d$) are
taken as unknowns, many of the electroweak observables such as
$A_{LR},A_{FB}^b,\Gamma_Z,\Gamma_{\bar\nu\nu},$ etc.\ become
independent of each other.  The $U(1)$ charges may even depend on the
fermion generation, as far as the constraints on flavor-changing
neutral currents are respected.  It is interesting that the numerical
quality of the present electroweak fit is rather poor~\cite{EWprec},
so there might already be a hint of such new-physics contributions.
On the other hand, for a specific model with fixed $U(1)$ charge
assignments, it is straightforward to include the appropriate
modifications in the expressions for the $Z$-fermion couplings.  The
remaining free parameters are formally equivalent to the $S$ and $T$
parameters which we have discussed in the preceding section.  However,
the numerical fit to the electroweak data has to be reconsidered in
this framework~\cite{Csaki2}.

\subsection{Four-Fermion Interactions}

At very low energies, the $W$ and $Z$ bosons are also integrated out
and give way to the four-fermion interactions of the Fermi model.
These interactions get corrections from the exchange of heavy vector
bosons, i.e., from the operators
\begin{equation}
  \Op^{(3)}_{JJ} = \tr{J^{(3),\mu} J^{(3)}_\mu}
\quad\text{and}\quad
  \Op^{(1)}_{JJ} = J^{(1),\mu} J^{(1)}_\mu.
\end{equation}
which are both present in (\ref{L6}).  Looking at charged-current
interactions, together with the shift in the $W$ mass this correction
effectively modifies the relation of the Fermi constant and the Higgs
vacuum expectation value~$v$:
\begin{equation}\label{z}
  \sqrt2\,G_F = \frac{1}{v^2}(1 + z)
\quad\text{with}\quad
  z = -\alpha\Delta T -\frac{v^2}{4} f^{(3)}_{JJ}.
\end{equation}

It is customary to choose $G_F$ (as measured in muon decay) as an
independent parameter of the SM.  If this is complemented by $M_Z$ and
$\alpha$ (i.e., $e$), we have to account for the shifts in the vector
boson masses,
and define the parameters $\vv$ and $\sw$ by the relations
\begin{equation}
  \vv = (\sqrt2\,G_F)^{-1/2}
\quad\text{and}\quad
  M_Z = \frac{e\vv}{2\sw\cw}.
\end{equation}
The two definitions of the weak mixing angle are thus related by
\begin{align}
  s_w^2 &= \sw^2 \left(1 + \frac{\cw^2}{\cw^2-\sw^2}(y+z)\right),
\notag \\ 
  c_w^2 &= \cw^2 \left(1 - \frac{\sw^2}{\cw^2-\sw^2}(y+z)\right).
\end{align}

\subsection{Constraints on the Littlest Higgs Model}

Electroweak precision data constrain the allowed parameter space of
Little Higgs models.  We have seen that in the case of universal
hypercharges, up to the order $v^2/F^2$, all corrections to low-energy
observables can be parameterized in terms of $\Delta S$, $\Delta T$,
and two extra parameters which introduce contact interactions of the
triplet and singlet currents.  Even in the non-universal case, for any
specific model where all hypercharge assignments are fixed we may take
into account their effects in the electroweak observables explicitly.
Then, in addition to $S$ and $T$, the coefficient $f_{VN}^{(1)}$ is left as
a free parameter, which can be constrained by the analysis of $Z$
decays.

Contact interactions have been sought for both at hadron and lepton
colliders.  Since they are formally of higher order on the $Z$ pole
(the interference of signal and background vanishes on the resonance),
they yield an independent set of constraints.  The exact form depends
on the $U(1)$ charge assignments.

For illustration, let us consider the original version of the Littlest
Higgs model~\cite{LHmin} where the situation is particularly simple,
since all fermions couple to the first $U(1)$ group only ($a=0$ and
$J_N=0$).  The present exclusion limits for $Z'$ bosons~\cite{Zprime}
can be turned into limits on the values of $f^{(3)}_{JJ}$ and
$f^{(1)}_{JJ}$ (\ref{f3jj}, \ref{f1jj}), i.e., on the ratios $c^2/F$
and $c^\pp/F$.  Thus, for a given value of $F$ these constraints can
be evaded if $c$ and $c'$ are both small.  This is the limit where the
massive vector bosons become superheavy and simultaneously decouple
from fermions.  In fact, for $c$ or $c'$ less than about $0.1$, the
vector boson masses are of the same order as the cutoff $\Lambda$
where the Little Higgs model breaks down as a low-energy effective
theory, and new (strong) interactions may be expected.

From current experimental data, the combined limit for a $Z'$ boson
with SM-like couplings is $M_{Z'}~\gtrsim~1.5\;\TeV$~\cite{Zprime}.
For the Littlest Higgs model, this translates roughly into
\begin{equation}
  c^2 \lesssim F/4.5\;\TeV
\quad\text{and}\quad
  c^\pp \lesssim F/10\;\TeV.
\end{equation}
The limits for charged heavy vector bosons are somewhat weaker.

In the limit $c\sim c'\to 0$ where all contact interactions disappear,
the correction to $S$~(\ref{deltaS}) also vanishes.  However, there
remains a constant contribution to $T$~\cite{Han},
\begin{equation}\label{dT0}
  \alpha\Delta T(c'=0) = 
  \frac{5v^2}{4F^2} - \frac{2v^2\lambda_{2\phi}^2}{M_\phi^4},
\end{equation}
where the second term depends on the parameters in the Coleman-Weinberg
potential.

The first term in~(\ref{dT0}) is due to the existence of the heavy
hypercharge boson~$Y$.  This positive shift in $T$ pushes the model
out of the exclusion contour in the $S$-$T$ plane allowed by
electroweak data for a light Higgs boson, unless $F$ is larger than
about $4\;\TeV$.  The second term, the shift due to heavy-scalar
exchange, is negative.  However, the bound~(\ref{lambda-min}) implies
that the net $\Delta T$ will not be smaller than $v^2/\alpha F^2$.

When discussing Little Higgs models, it is usually assumed that the
Higgs boson is light, presumably close to the lower experimental limit
$m_0\approx 115\;\GeV$.  However, this is not necessarily true:
Depending on the parameters in the Coleman-Weinberg potential (e.g.,
if the two contributions in the denominator of~(\ref{Hmass}) almost
cancel each other), the physical Higgs mass can take any value that is
not in conflict with unitarity.
Increasing the Higgs mass with respect to the reference value $m_0$, we
get additional shifts in $S$ and $T$.  These are approximately given
by~\cite{STU}
\begin{equation}
  \Delta S = \frac{1}{12\pi}\ln\frac{m_H^2}{m_0^2}
\quad\text{and}\quad
  \Delta T = -\frac{3}{16\pi c_w^2}\ln\frac{m_H^2}{m_0^2}
\end{equation}
(The complete one-loop formulae can be found, e.g., in~\cite{Zep}.)
As a consequence, the positive contribution to $T$ can be partially
cancelled by an increase in the Higgs mass.  Note that this would also
reduce the amount of fine-tuning in the model.

In the presence of oblique {\em and} non-oblique corrections one has
to carefully define the $S$ and $T$ parameters.  In our effective
theory they are defined on the operator level and can be read off from the
gauge boson masses~(\ref{mw-mz}), if the scale $v$ is given.
Introducing the abbreviations $g_W = e/s_w$ and $g_Z = e/(s_w c_w)$
for the couplings of $W$ and $Z$ to fermions ($e$ and $s_w$ defined at
the $Z$ pole), using (\ref{geff}, \ref{gzeff}, \ref{mw-mz}--\ref{y})
we can make up dimensionless ratios
\begin{align}
  \frac{M_W^2}{c_w^2 M_Z^2} &=
	 1 - \frac{\alpha}{4c_w^2} \Delta S + \alpha \Delta T, \label{MW/MZ}\\
  \frac{\Gamma_W^2}{M_W^2} &\propto 
         g_W^4 (1 + 2 \frac{\alpha}{4s_w^2} \Delta  S), \label{GW/MW} \\
  \frac{\Gamma_Z^2}{M_Z^2} &\propto 
         g_Z^4 (1 +  2 \frac{\alpha}{4s_w^2c_w^2}\Delta S ) \label{GZ/MZ},
\end{align}
where $v$ drops out.  Here, $\Gamma_{W/Z}$ stand for either the total
width or for a partial decay width of the corresponding vector boson.
The prefactors in (\ref{GW/MW}, \ref{GZ/MZ}) are known functions
which at leading order just depend on $s_w$, while higher-order
corrections, in a consistent approximation, add incoherently to the
new-physics contribution considered here.  Accepting the fact that
$\Gamma_W$ is not sufficiently well measured to be relevant here, we
nevertheless can extract $S$ and $T$ from (\ref{MW/MZ}) and
(\ref{GZ/MZ}) alone, i.e., exclusively from $Z$- and $W$-pole data.

This is not the conventional way of extracting the oblique
parameters~\cite{EWprec,PPP03}, where $e$ and $s_w$ are used as above, but
the low-energy observable $G_F$ is included as a dimensionful quantity
which sets the scale~$v$.  Due to the presence of non-oblique new
physics, the relation between $G_F$ and $v$ is modified by an amount
$z$ according to (\ref{z}), which accounts for the shifts in the
electroweak couplings and the $W$ mass as well as a triplet contact
term.  Generically, for any model with a heavy gauge triplet we have
\begin{equation}
  z \equiv -\alpha \Delta T + \delta
\end{equation}
with $\delta = (c^2 v/F)^2$, where $c$ is the cosine of the mixing
angle between the two $SU(2)$ and $F$ the high scale. 

In the presence of non-oblique corrections we may call the conventional
definition of $S$, $T$, and $U$ {\em effective} parameters, which in
the linear approximation are given by~\cite{EWprec}
\begin{equation}
  \label{stu_pdg}
  \begin{aligned}
    \frac{M_W^2}{M_{W,0}^2} &= 
	1 + \frac{\alpha}{4s_w^2} (S_\eff +  U_\eff) ,\\
     \frac{M_Z^2}{M_{Z,0}^2} &= 
        1 + \frac{\alpha}{4s_w^2c_w^2} S_\eff - \alpha T_\eff,    \\
    \frac{\Gamma_Z}{M_Z^2 \beta_Z} &= 
        1 + \alpha T_\eff,
  \end{aligned}
\end{equation}
where $\beta_Z = \Gamma_{Z,0}/M_{Z,0}^3$.  The quantities with the
zero subscript are calculated in the pure SM, i.e, using the {\em
measured} values of $e$, $s_w$, and $G_F$.  While the extraction of
$e$ from electromagnetic data and $s_w$ from $Z$-pole asymmetries is
free of non-oblique corrections, $G_F$ contains an extra
contribution $z$~(\ref{z}) if related to the electroweak scale~$v$.

In our effective theory, we obtain 
\begin{equation}
  \begin{aligned}
     \frac{M_W^2}{M_{W,0}^2} &= 
       1 + \frac{\alpha}{4s_w^2} \Delta S + \delta ,\\
     \frac{M_Z^2}{M_{Z,0}^2} &= 
       1 + \frac{\alpha}{4s_w^2c_w^2} \Delta S - \alpha\Delta T + \delta,    \\
    \frac{\Gamma_Z}{M_Z^2 \beta_Z} &=
       1 + \alpha \Delta T - \delta.    
  \end{aligned}
\end{equation}
A comparison with (\ref{stu_pdg}) reveals the connection between the
effective parameters and the ones we have calculated above,
\begin{align}
  S_\eff &= \Delta S, \\
  T_\eff &= \Delta T - \frac{1}{\alpha} \delta,\\
  U_\eff &= \frac{4s_w^2}{\alpha} \delta. 
\end{align}
The result is somewhat unexpected: The choice of the low-energy
parameter $G_F$ as input mimics nonvanishing $T_\eff$ and $U_\eff$
even in the absence of custodial $SU(2)$ violation.  Actually,
low-energy neutral-current data or a precise measurement of the $W$
width would allow for the identification of the non-oblique correction
$\delta$, but the present experimental accuracy is insufficient for
this, given the small values of $\delta$ that are allowed by the
direct constraints on heavy vector bosons.

In the Littlest Higgs model, for increasing mixing angle $c$ the shift
in $T_\eff$ (which is $\propto c^4$) compensates the positive $\Delta T$
contribution of the $SU(2)_c$-violating sector, eventually resulting
in a negative $T_\eff$ value. 

\begin{figure*}
  \begin{center}\includegraphics{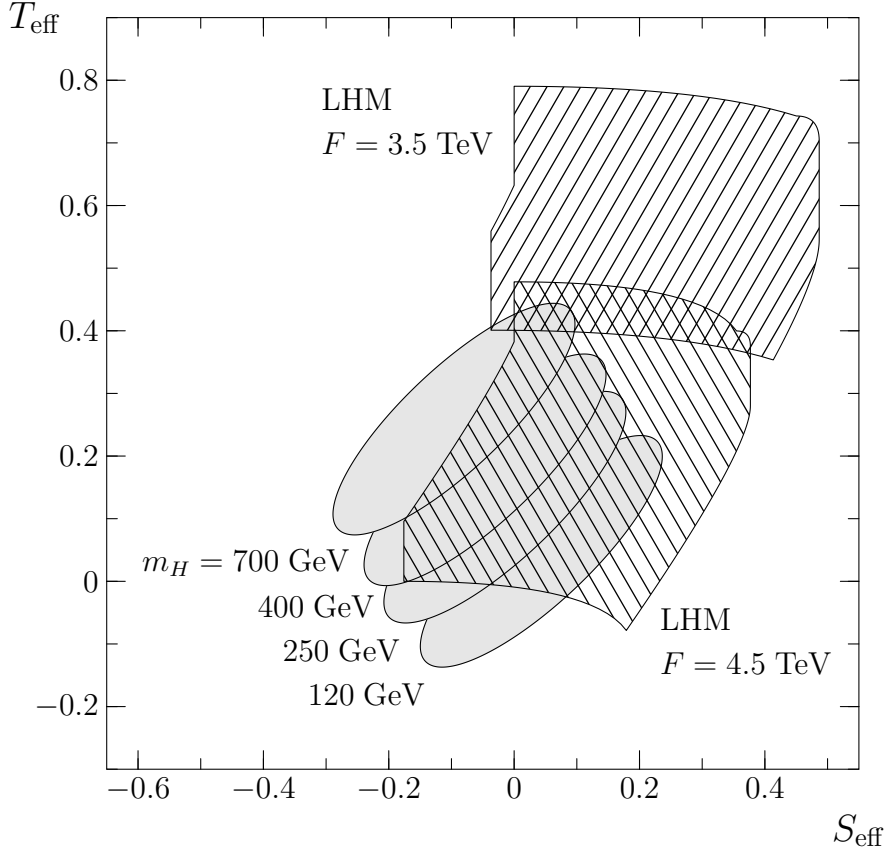}\end{center}
  \vspace{\baselineskip}
  \caption{$ST$ predictions for the Littlest Higgs model with standard
  $U(1)$ charge assignments~\cite{LHmin}.  The shaded
  ellipses are the $68\,\%$ exclusion contours which follow from the
  electroweak precision data~\cite{EWprec,PPP03}, assuming four
  different Higgs masses.  The hatched areas are the allowed parameter ranges
  of the Littlest Higgs model for two different values of the
  scale~$F$.  The limits from contact interactions have been taken
  into account.}
  \label{fig:ST}
\end{figure*}

In Fig.~\ref{fig:ST}, we depict the allowed region in the
$S_\eff$-$T_\eff$ plane for two different values of $F$.  The contours
are restricted by the direct limits on contact interactions.  After
having translated our $\Delta S$, $\Delta T$ parameters to the
effective ones, we can take the fit of the $ST_\eff$-parameters for a
given Higgs mass \emph{as is}~\cite{EWprec} and compare it with the
prediction of the model under consideration.  Looking at the
figure, we can conclude that a light Higgs boson ($m_H=120\;\GeV$) is
consistent with the Littlest Higgs model only if $F\gtrsim
4\;\TeV$~\cite{Csaki,Hewett,Han}.  However, allowing for larger Higgs
masses reduces this limit to less than $4\;\TeV$.  We should also keep
in mind that radiative corrections~\cite{rho-loop} and unknown effects
from new physics beyond the UV cutoff $\Lambda$ will add extra small
shifts to the $S$ and $T$ parameters which can slightly change this
conclusion.

Our derivation shows how this picture looks like in more general
models.  In the triplet sector, any extension will only result in
additional contributions which have a form identical to that in the
Littlest Higgs model.  These terms will contribute to $\Delta S$ only.  In
the singlet sector, there is more freedom: Removing the $U(1)$ boson,
changing the hypercharge assignments, or extending this sector in some
other way, allows for different values of $\Delta
T$~\cite{Han,Csaki2,PPP03}, and more free parameters may enter the
game.  In particular, the spectrum can be arranged to be consistent
with the custodial $SU(2)$ symmetry, such that the corrections to the
$T$ parameter cancel altogether.  In such models, the bounds on $F$
are significantly weaker than in the Littlest Higgs
model~\cite{LHmod}.

\section{New Effects}

New data from Tevatron, LHC, and a future Linear Collider will allow
us to constrain the parameter space by measurements of new independent
observables.  Apart from improved limits on contact interactions,
there will be precision data on vector-boson self-couplings, on Higgs
and top-quark interactions.  In this section, we derive the
corresponding anomalous contributions, starting
from the effective Lagrangian~(\ref{L6}).  

The results given below apply directly to any Little Higgs model which
contains no extra singlet current and only one light Higgs doublet.
The complications which arise in the presence of an extra current have
been discussed above.  In addition to the anomalous $Z$ couplings, the
operator $\Op_{VN}^{(1)}$~(\ref{op1vn}) induces $HZff$ interactions which
add to the terms in~(\ref{L-HVJ}).  As far as the Higgs sector is
concerned, many Little Higgs models predict more than one light Higgs
doublet.  However, in the present paper we do not attempt a discussion
of extended Higgs sectors, and leave the general case to a future
publication.  If there are multiple physical Higgs states, the
structure of anomalous couplings in the gauge and fermion sectors is
unchanged, but the genuine Higgs-gauge, Higgs-fermion and Higgs
self-couplings reflect the additional complications.

\mbox{}

\subsection{Anomalous Triple-Gauge Couplings}

Strictly speaking, tree-level contributions to triple gauge couplings
from the exchange of heavy particles are impossible at the level of
dimension-six operators~\cite{Wudka94}.  In the absence of fermions
the self-couplings of vector bosons define the gauge couplings $g$ and
$g'$.  However, in practice the gauge couplings are measured in
processes which involve external fermions.  These interactions do
receive tree-level corrections.  We have used the equations of motion
to canonically normalize the fermion gauge interactions.  As a result,
anomalous triple-gauge couplings appear.

We use the standard parameterization:
\begin{widetext}
\begin{equation}
\begin{split}
  \label{anomalous}
  \LL_{WWV} &= 
    -\ii e\frac{c_w}{s_w} \Bigl[ 
    g_1^Z \left( W^{+,\,\mu\nu} W_\mu^- - W^+_\mu W^{-,\,\mu\nu} \right) Z_\nu 
    + \kappa_Z W_\mu^+ W_\nu^- Z^{\mu\nu} \Bigr] 
  \\ &\quad
    -\ii e \Bigl[ 
    g_1^\gamma
      \left( W^{+,\,\mu\nu} W_\mu^- - W^+_\mu W^{-,\,\mu\nu} \right)  A_\nu 
    + \kappa_\gamma W_\mu^+ W_\nu^- A^{\mu\nu} \Bigr] 
\end{split}
\end{equation}
\end{widetext}
Inserting our results into the formulae of~\cite{Zep}, we obtain
\begin{equation}
  g_1^Z = \kappa_Z = 1 - 2M_Z^2 f_{VW}
\quad\text{and}\quad
  g_1^\gamma = \kappa_\gamma = 1.
\end{equation}
Since the anomalous contributions are generated solely by a
renormalization of gauge couplings, we have $g_1^Z=\kappa_Z$ and
$g_1^\gamma=\kappa_\gamma$.  Electromagnetic gauge invariance requires
$g_1^\gamma=1$, hence the photon couplings are unchanged.  The
$SU(2)_c$ relation
\begin{equation}
  \Delta\kappa_\gamma = -\frac{c_w^2}{s_w^2}(\Delta\kappa_Z-\Delta g_1^Z)
\end{equation}
is automatically satisfied in this case, so there is no contribution
from the $SU(2)_c$-violating coefficient $f_{VB}$.

In our derivation of the effective Lagrangian, we have expressed
everything in terms of the parameters $e$, $s_w$, and a dimensionful
quantity, for which we either take $M_W$, $M_Z$, or $v$.  As far as
the coefficients of dimension-six operators are concerned, the
particular scheme used for experimentally defining the input
parameters is irrelevant, since differences are formally of higher
order.  However, when discussing interactions which also occur at tree
level in the Standard Model, we have to be specific about the
definition of the input parameters, since a shift in the SM
contribution will be of the same order as the direct contribution of
the anomalous interactions.

For instance, in the $G_F$-$M_Z$-$\alpha$ scheme, there is an
additional contribution to $g_1^Z$ and $\kappa_Z$ since we have to
define the three-gauge-boson vertices in terms of $\sw$ and $\cw$:
\begin{multline}
  \LL_{WWZ} = 
    -\ii e\frac{\cw}{\sw} \Bigl[ 
    g_1^Z \left( W^{+,\,\mu\nu} W_\mu^- - W^+_\mu W^{-,\,\mu\nu} \right) Z_\nu 
    \\ + \kappa_Z W_\mu^+ W_\nu^- Z^{\mu\nu} \Bigr],
\end{multline}
where now
\begin{equation}
  g_1^Z = \kappa_Z = 1 - 2M_Z^2f_{VW} - \frac{y+z}{2(\cw^2-\sw^2)}.
\end{equation}

\subsection{Anomalous Higgs couplings}
\label{sec:anhiggs}

Anomalous Higgs couplings are induced both by vector and scalar
exchange.  Expanding the effective Lagrangian~(\ref{L6}), we obtain the
following contributions:
\begin{enumerate}
\item
  Couplings to longitudinal gauge bosons:
  \begin{equation}\label{HVV}
    \LL_{HVV} = \chi_{W}\frac{2M_W^2}{v}HW^{+,\mu}W^-_\mu 
              + \chi_{Z}\frac{M_Z^2}{v}HZ^\mu Z_\mu
  \end{equation}
  These occur at tree-level in the SM and get anomalous contributions
  from various sources.  The direct contributions are
  \begin{align}
    \chi_{W} &= 1 + 4M_W^2 f_{VW} 
    + \frac{v^2}{2}f_{hh},
    \\
    \chi_{Z} &= 1 + 4\left(M_W^2 f_{VW} + 2s_w^2 M_Z^2 f_{VB}\right)
    \notag \\ & \quad + \frac{v^2}{2}\left(f_{h,1} + f_{hh}\right).
  \end{align}
  These relations apply if we define the parameters $M_W$ and $M_Z$ in
  (\ref{HVV}) as the measured values (\ref{mw-mz}).  Due to the
  appearance of $v$ in (\ref{HVV}), in the $G_F$-$M_Z$-$\alpha$
  scheme there is also an indirect contribution to be added:
  \begin{equation}
    \Delta \chi_{W} = \Delta \chi_{Z} = -\frac12 z,
  \end{equation}
  where $z$ is given in (\ref{z}).
\item
  Couplings to transversal gauge bosons:
  \begin{align}
    \LL_{HVV}' &= h'_{WW}HW^{+,\mu\nu}W^-_{\mu\nu}
       + h'_{ZZ}HZ^{\mu\nu}Z_{\mu\nu} \notag \\ & \quad
       + h'_{ZA}HZ^{\mu\nu}A_{\mu\nu}
       + h'_{AA}HA^{\mu\nu}A_{\mu\nu}
  \end{align}
  While these terms are generated by the operators $\Op_{BW}$,
  $\Op_{WW}'$, and $\Op_{BB}'$ individually, they all vanish in the linear
  combination present in (\ref{L6}), and thus are induced at loop-level
  only, as one would expect.
\item
  Contact terms, i.e., direct couplings of Higgs bosons to vector
  bosons and the weak or hypercharge currents.  These follow from the
  (\ref{L6}), if the physical fields are inserted, derivatives acting
  on the Higgs field are eliminated by partial integration, and the
  equations of motion of the vector fields are applied.
  Alternatively, we can read them off directly from the $V$-$J$
  interactions in (\ref{L-3},\ref{L-1}):
  \begin{equation}
  \begin{split}\label{L-HVJ}
    \LL_{HVJ} &=
     M_W f^{(3)}_{VJ} \left(HW^{+,\mu}J^-_\mu + HW^{-,\mu}J^+_\mu\right)
    \\ &\quad
     + M_Z f^{(3)}_{VJ} HZ^\mu J^{(3),0}_\mu
    \\ &\quad
     - 2M_Z HZ^\mu \left(f^{(1)}_{VJ} J^{(1)}_{Y,\mu} 
                         + f^{(1)}_{VN} J^{(1)}_{N,\mu}\right). 
  \end{split}
  \end{equation}
\item
  Anomalous couplings of the Higgs boson to the scalar current, i.e.,
  to massive fermions.  Due to the effect of the operator
  $\Op_{J_S}'$, all such couplings are modified by the common factor
  \begin{equation}\label{chif}
    \chi_f = 1 - v^2 f_{J_S} 
    = 1 + \frac{v^2}{2}\left(f_{VV}^{(3)} + 2 f_{VV}^{(1)}\right).
  \end{equation}
  If the fermions are mixed with new heavy particles, as it is the
  case for the top quark in Little Higgs models, there are extra
  contributions to~(\ref{chif}).  These will be considered below in
  Sec.~\ref{sec:tcoup}.
\end{enumerate}

\subsection{Higgs pairs}

In the effective Lagrangian~(\ref{L6}), various terms induce anomalous
couplings which are relevant for Higgs pair production.
\begin{enumerate}
\item
  The quartic $HHWW$ and $HHZZ$ couplings are modified:
  \begin{equation}
    \LL_{HHVV} =
    \eta_{W}\frac{M_W^2}{v^2} H^2 W^{+,\mu} W^-_\mu
    + \eta_{Z}\frac{M_Z^2}{2v^2} H^2 Z^{\mu} Z_\mu,
  \end{equation}
  where
  \begin{align}
    \eta_{W} &= 1 + 20M_W^2 f_{VW} 
    -\frac{v^2}{2}\left( f_{h,1} - 4f_{hh}\right),
  \\
    \eta_{Z} &= 1 + 20\left(M_W^2 f_{VW} + 2s_w^2M_Z^2f_{VB}\right)
    \notag \\ & \quad + 2v^2\left( f_{h,1} + f_{hh}\right).
  \end{align}
  are the direct contributions.  Here, we have applied the equations
  of motion of the Higgs boson to eliminate derivative couplings.  The
  indirect corrections in the $G_F$-$M_Z$-$\alpha$ scheme are
  \begin{equation}
    \Delta\eta_{W} = \Delta\eta_{Z} = - z.
  \end{equation}
\item
  The cubic Higgs self-coupling is directly affected by the presence
  of the operator $\Op_{h,3}$.  Furthermore, the operators $\Op_{h,1}$
  and $\Op_{hh}$ contribute to this coupling if we eliminate
  derivative couplings by the equations of motion.  Parameterizing the
  vertex by
  \begin{equation}
    \LL_{HHH} = - \chi_H\frac{m_H^2}{2v}H^3,
  \end{equation}
  we have a direct contribution
  \begin{align}
    \chi_H &= 
    1 - \frac{v^2}{2}\left(f_{h,1} + f_{hh}\right)
    - \frac{2v^4}{3m_H^2}f_{h,3}.
  \end{align}
  To determine
  the indirect contribution, we augment the set of independent
  parameters ($G_F$, $M_Z$, $\alpha$) by the physical Higgs mass
  $m_H$, to get
  \begin{equation}
    \Delta\chi_H = -\frac12z
  \end{equation}
  in this scheme.
\end{enumerate}

\subsection{Top-Quark Couplings}
\label{sec:tcoup}

The presence of heavy vector-like quarks in the spectrum affects
the interactions of the top quark with gauge bosons and Higgs bosons
in a non-universal way.
\begin{enumerate}
\item
  The electroweak interactions of the top and bottom quarks are
  modified by the operators $\Op_{Vq}$ and $\Op_{Vt}$~(\ref{opvq},
  \ref{opvt}), which originate from heavy quark exchange, and by the
  redefinition of the vector fields due to heavy vector exchange.  In
  the physical basis, the $Z$ and $W$ couplings are
  \begin{equation}
  \begin{split}
    \LL_{tV} &= -\frac{e}{4c_ws_w}
      \left[
        \bar t \fmslash{Z} \left(v_t - a_t\gamma_5\right) t 
        -\bar b \fmslash{Z} \left(v_b - a_b\gamma_5\right) b
      \right]
  \\ &\quad
      - \frac{e}{2\sqrt2\,s_w} c_{tb}
        \left[\bar t \fmslash{W}^+(1-\gamma_5)b 
              + \bar b \fmslash{W}^-(1-\gamma_5)t\right]
  \end{split}
  \end{equation}
  where the coefficients are given by
  \begin{align}
    v_b &= \left(1-\frac43s_w^2\right)
      \left[1 + {M_Z^2}\left(f_{VW}+2f_{VB}\right)\right]
  \\
    a_b &= 1 + {M_Z^2}\left(f_{VW}+2f_{VB}\right)
  \\
    v_t &= \left(1-\frac83s_w^2\right)
      \left[1 + {M_Z^2}\left(f_{VW}+2f_{VB}\right)\right]
\notag\\ &\quad
      + v^2 (f_{Vq} + f_{Vt})
  \\
    a_t &= 1 + {M_Z^2}\left(f_{VW}+2f_{VB}\right) + v^2 (f_{Vq} - f_{Vt})
  \\
    c_{tb} &= 1 + {M_W^2}\left(f_{VW}+2f_{VB}\right) 
                + \frac{v^2}{2} f_{Vq}
  \end{align}
  While the corrections proportional to $(f_{VW}+2f_{VB})$ are
  universal to all fermions and taken into account by the $S$-$T$ fit
  of the SM, the corrections proportional to $f_{Vq}$ and $f_{Vt}$ are
  specific to the top-quark vertices.

  The indirect corrections in the $G_F$-$M_Z$-$\alpha$ scheme are in
  this case
  \begin{align}
    \Delta v_b &= -
      \left(1 + \frac{4\sw^2/3}{\cw^2-\sw^2}\right)\frac{y + z}{2},
  \\
    \Delta v_t &= -
      \left(1 + \frac{8\sw^2/3}{\cw^2-\sw^2}\right)\frac{y + z}{2},
  \\
    \Delta a_b &= \Delta a_t = - \frac{y + z}{2},
  \\
    \Delta c_{tb} &= -\frac{\cw^2}{\cw^2-\sw^2}\,\frac{y + z}{2}.
  \end{align}

\item
  The top-quark Yukawa coupling is also modified by the heavy
  $T$-quark.  There are further effects due to nonlinear Goldstone
  interactions and heavy-scalar exchange which altogether make up the
  coefficient of the operator $\Op_{hq}$.  Finally, there are the
  corrections from $\Op_{VV}^{(3)}$ and $\Op_{VV}^{(1)}$ which have
  been given already in~(\ref{chif}).  The resulting vertex is
  \begin{equation}
    \LL_{tH} = -\frac{m_t}{v}\chi_t\bar t H t,
  \end{equation}
  where
  \begin{equation}
    \chi_t = 1 + \frac{v^2}{2}\left(f_{VV}^{(3)} + 2f_{VV}^{(1)}\right)
               + \frac{v^3}{\sqrt2\,m_t} f_{hq}.
  \end{equation}
  The indirect contribution in the $G_F$-$M_Z$-$\alpha$ scheme is
  \begin{equation}
    \Delta\chi_t = -\frac12 z.
  \end{equation}
\item
  There are also quartic $ttZH$ and $tbWH$ vertices:
  \begin{equation}
  \begin{split}
    \LL_{tVH} &= 
      -M_Z f_{Vq} \bar t H\fmslash{Z} \left(1 - \gamma_5\right) t 
  \\ &\quad
      -M_Z f_{Vt} \bar t H\fmslash{Z} \left(1 + \gamma_5\right) t 
  \\ &\quad
      - \frac{1}{\sqrt2} M_W f_{Vq}
        \Bigl[\bar t H\fmslash{W}^+(1-\gamma_5)b 
  \\ &\qquad\quad  
  + \bar b H\fmslash{W}^-(1-\gamma_5)t\Bigr]
  \end{split}
  \end{equation}
  In the Littlest Higgs model, $f_{Vt}=0$, and these couplings are
  purely left-handed, which is due to the fact that it is the
  right-handed top quark which mixes with the heavy $T$ fermion.
\end{enumerate}

\section{The Reconstruction of a Little Higgs Model}

Despite the fact that Little Higgs Models are constrained by
electroweak precision data, there remains a considerable parameter
space where such models are viable.  Assuming that such a mechanism is
realized in nature, one should ask the question to what extent it is
possible to derive the model and its parameters from experiments at
future colliders.

To verify the generic mechanism that is common to all Little Higgs
models, we would like to check two characteristic properties of the
model, namely the cancellation of quadratic divergences as a result
of the symmetry structure, and the Goldstone-boson nature of the Higgs
boson.  A direct check of the first property would require the
measurement of the quartic couplings of Higgs bosons to heavy vectors,
scalars, and fermions, which is out of reach of the next generation of
colliders.  However, the symmetry structure manifests itself also in
relations of couplings which are accessible once the new particles
have been discovered at the LHC~\cite{Han,PPP03,BPP03}.  The same
couplings also enter the low-energy effective Lagrangian.  For
instance, the cancellation in the vector-boson triplet sector is
reflected in the relation of the coefficients $f_{JJ}^{(3)}$ and
$f_{VW}$ (\ref{f3jj}, \ref{fvw}), once the scale $F$ is known.
Similar statements hold for the scalar and fermion sectors.  Thus, a
sufficiently accurate determination of the low-energy coefficients
complements direct measurements at LHC.  In cases where direct
measurements are difficult (e.g., in the scalar sector), low-energy
observables may be the only handle on the Little Higgs mechanism.

In order to establish the Goldstone nature of the Higgs boson, we
should demonstrate the nonlinearity in the Higgs-boson representation
above the scale $F$, i.e., the presence of nonrenormalizable terms in
the Higgs interactions.  The low-energy trace of this is encoded in
terms which are independent of the mixing angles and masses of the
Little Higgs spectrum.  For instance, in the Littlest Higgs model
there are a constant contribution $-1/6F^2$ in the coefficient
$f^{(3)}_{VV}$~(\ref{f3vv}) and a similar term in the coefficient
$f_{hq}$~(\ref{fq}) (i.e., the constant $2/3$ in
$\beta$~(\ref{beta})).

Since the anomalous contributions we have calculated in the preceding
sections all carry a common suppression factor $v^2/F^2$ relative to
the SM result, for a meaningful measurement the low-energy observables
have to be determined at least to this accuracy.  If $F$ happens to be
rather high (e.g., $F\gtrsim 4\;\TeV$ for the unmodified Littlest
Higgs model), the required precision is in the per mil range.  A
high-luminosity $e^+e^-$ Linear Collider can reach this level for a
limited subset of observables which include contact terms and
triple-gauge couplings.  In the Higgs and top sectors, accuracies of
the order of $1$-$2\,\%$ are possible for the observables of
interest~\cite{TESLA}.  At LHC, the level of precision is generically
weaker, but direct measurements are possible for new heavy particles
in the spectrum.  Thus, if the scale $F$ is of the order $2\;\TeV$ or
less, which is allowed in various Little Higgs
models~\cite{Csaki2,LHmod}, a complete coverage of the
low-energy parameters becomes feasible.  In any case, all observables
will be included in a combined fit if signals of a Little Higgs model
are found, once a sufficient data sample has been collected at LHC and
a Linear Collider.

\subsection{Vector Bosons}

The new $X$ and $Y$ gauge bosons can be produced and detected at the
LHC if they are not too heavy~\cite{Han}, and their couplings can be
directly measured~\cite{BPP03}.  Indirect constraints from low-energy
observables can be combined with those results for an improved fit and
will help to disentangle the contributions of various sectors:
\begin{enumerate}
\item
  The measurement of contact terms, e.g., in the processes
  \begin{equation}
    e^+e^- \to e^+e^-
  \quad\text{and}\quad
    e^+e^- \to \mu^+\mu^-
  \end{equation}
  will significantly improve the limits for a particular combination
  of the operator coefficients $f^{(1)}_{JJ}$ and $f^{(3)}_{JJ}$,
  equivalent to the detection of a $Z'$ boson up to a mass of
  $5$-$10\;\TeV$~\cite{ZprimeT,TESLA}.  For a separate measurement of the
  triplet contribution, one needs a charged-current channel.  For
  instance, the cross section measurement of the process
  \begin{equation}
    e^+e^- \to \bar\nu\nu\gamma
  \end{equation}
  allows for detecting the effect of $W'$ bosons up to $M\sim
  5\;\TeV$~\cite{Wprime,TESLA}.  These limits can be combined with the
  possible observation (or non-observation) of those states at LHC to
  extract the scale $F$ and the mixing angles in the vector-boson sector.
\item
  Another probe of heavy-vector exchange is given by quartic $HZff$
  and $HWff$ interactions, which depend on $f^{(3)}_{VJ}$ and
  $f^{(1)}_{VJ}$.  The neutral component can be extracted by measuring
  the angular distribution and/or the energy dependence of the
  Higgs-strahlung process~\cite{KKZ95}, while the charged component
  affects $WW$ fusion.  (A detailed experimental analysis of contact
  terms in Higgs production has not yet been performed.)
\item
  The triple-gauge couplings will be measured to better than per mil
  accuracy at a Linear Collider~\cite{TGVexp,TESLA}.  Assuming that
  $S$ and $T$ and the contact terms are known, this allows for the
  extraction of the coefficient $f^{(3)}_{VJ}$ to a precision level
  comparable to the contact-term measurements.  Thus, an independent
  check of the coupling relations in the vector-boson sector is
  feasible.
\item
  Once the Higgs mass is known, the existing precision data can be
  turned into measurements of $\Delta S$ and $\Delta T$.  If the
  Giga-$Z$ option of a Linear Collider is realized, the accuracy of
  this measurement will improve by one order of
  magnitude~\cite{TESLA}.  In our context, the value of $\Delta S$
  provides us with the parameter combination $f_{VW}+2f_{VB}$.
  Turning the argument around, together with the measurement of triple
  gauge couplings one gets an independent constraint on
  $f^{(3)}_{JJ}$.
\end{enumerate}

Combining those measurements in a single fit, all parameters in the
gauge sector can be derived.  In particular, if LHC and Linear
Collider data are taken together, there will be enough redundancy to
go beyond the assumption of a specific model, such that the complete
set of heavy vector bosons (singlets and triplets) and their couplings
can be reconstructed.

\subsection{Scalars}

In the effective Lagrangian~(\ref{L6}), Higgs-boson operators are
affected both by the scalar and by the vector boson sector.  Since the
vector-boson contributions can be extracted by the methods described
above, we get an indirect handle on the scalar sector, which is
difficult to access directly.  The statistical and
systematic uncertainties for Higgs production and decay measurements
at LHC and a Linear Collider limit the achievable accuracy to
$1$-$2\,\%$ or worse, depending on the channel and on the Higgs
mass~\cite{TESLA,HanZ}.  The following arguments show that a complete
coverage of the scalar sector is possible in principle.  In practice,
this exercise can be successful if the scale $F$ is of the order
$2\;\TeV$ or lower, while for higher scales the accessible information
becomes limited.
\begin{enumerate}
\item
  $\Delta T$ depends on $f^{(1)}_{VV}$ and a correction due to
  heavy-scalar exchange.  Once the Higgs mass and the properties of
  new $U(1)$ vector bosons are known, we can isolate this piece.  This
  quantity (i.e., the $\rho$ parameter) has been measured with per mil
  accuracy.  At GigaZ this can be improved by another order of
  magnitude.
\item
  The couplings of the Higgs boson to gauge bosons will be measured in
  Higgs-strahlung and vector boson fusion.  Combining this with the
  information on $\Delta T$, we can constrain the coefficient
  $f^{(3)}_{VV}$.
\item
  The ratio of the branching ratios $H\to ff$ and $H\to WW,ZZ$ also
  depends on the coefficients $f^{(3)}_{VV}$ and $f^{(1)}_{VV}$.  Thus,
  Higgs decay measurements will add independent information on those
  coefficients.  (Here, we need the assumption that the fermions are
  not mixed with any heavy partners, which is likely true for the $b$
  quark, and even more for $\tau$ and~$c$.)
\item
  Finally, double Higgs production depends on the coefficient
  $f_{h,3}$, the Higgs potential correction.  Both at a Linear
  Collider and at LHC this measurement is severely
  statistics-limited~\cite{Gay,TESLA,Baur}, and in
  Little Higgs models the small corrections to the trilinear Higgs
  coupling are unobservable even for very low~$F$.
\end{enumerate}

If sufficient precision can be reached from a combination of all
available data, we can isolate the contribution of the heavy scalar
$\phi$ (and thus confirm its existence) and detect the constant
contribution in $f^{(3)}_{VV}$ (\ref{f3vv}) which stems from the last
term of~(\ref{L0G}).  As discussed above, this would be direct
evidence for the Goldstone-boson nature of the Higgs boson.

Our discussion has been centered on the Littlest Higgs model with its
obvious generalizations, which contains just a single Higgs doublet in
its low-energy spectrum.  Other Little Higgs models predict a richer
structure: Apart from extra doublets, there could also be light scalar
singlets and triplets, which have to be pair-produced and thus are
difficult to access.  While this complication will not invalidate our
treatment of the vector-boson sector, the reconstruction of the scalar
sector in such models is beyond the scope of the present paper.

\subsection{Top-Quark Observables}

The top quark will be studied both at LHC and at a Linear Collider.
In addition, LHC opens the opportunity to produce new states in the
quark sector directly (e.g., the heavy quark $T$ of the Littlest Higgs
model) and study their decays~\cite{Han,PPP03}.  Here, we consider the
information on this sector which low-energy observables can provide.
\begin{enumerate}
\item
  While $\bar t t$ production at threshold is dominated by QCD
  effects, continuum production of top pairs allow for an accurate
  determination of the form factors $v_t$ and $a_t$ and thus provides
  a measurement of the operator coefficients $f_{Vq}$ and $f_{Vt}$.
  The achievable accuracy is of the order
  $1$-$2\,\%$~\cite{top,TESLA}.
\item
  The same coefficients are probed by
  measurements of the $tbW$ vertex in single-top production and in top
  decays.
\item
  A measurement of the top Yukawa coupling (or the ratio
  $g_{ttH}/g_{bbH}$) complements this by information on the scalar
  couplings to the top sector, i.e., the coefficient $f_{hq}$.
  Similar to $f_{VV}^{(3)}$, this anomalous coupling contains a
  constant contribution which is not due to heavy-particle exchange,
  but a consequence of the nonlinear Goldstone nature of the Higgs
  boson.  Here, a Linear Collider could reach a precision of up to
  $2.5\,\%$ (depending on the Higgs mass)~\cite{TESLA}.
\end{enumerate}

The sensitivity to the top-quark sector of Little Higgs models from
low-energy observables is similar to the scalar sector.  However, the
prospects for direct measurements at the LHC are more favorable in
this case due to the fact that new heavy quarks can be produced by
strong interactions.

\section{Conclusions}

Using the effective-theory formalism, we have given a complete
account of the anomalous couplings that are present in Little Higgs
models below the new-particle threshold.  In models without extra
gauged $U(1)$ groups, all present-day constraints on the parameter
space are encoded in the parameters $S$ and $T$ and the coefficients
of four-fermion contact interactions.  If new $U(1)$ vector bosons
with non-canonical hypercharge assignments are present, they
complicate this picture and introduce shifts in the individual
$Z$-fermion couplings, so the electroweak fit has to be adapted
accordingly.

The existing constraints on $S$ and $T$, if combined with the limits
on contact interactions, push the expected scale $F$ of the minimal
Little Higgs model (the Littlest Higgs model in its original version)
up to $4\;\TeV$ and higher, where for a scale less than about
$5\;\TeV$ a high Higgs mass is necessary to fit the electroweak data.
This lower limit is mainly caused by the large amount of custodial
$SU(2)$ violation in this model, and it can be evaded by a different
treatment of the hypercharge sector.  The limits on the $S$ parameter
are similarly constraining and restrict the mixing angles in the
vector-boson sector.

Beyond the electroweak precision observables which have been measured
so far, new anomalous couplings exhibit traces of all sectors of the
model.  While some masses and couplings can be determined directly in
the production and decay of heavy particles at the LHC, it becomes
possible to derive the full structure of the model by combining this
with precision measurements at a future Linear Collider.  If the achievable
precision is sufficient (this strongly depends on the actual value of
the new-physics scale $F$), we will be able to check the coupling
relations that are responsible for the cancellation of divergences and
the nonlinear nature of the Higgs interactions, thus verifying or
excluding the Little Higgs mechanism as the correct model of
electroweak symmetry breaking.

\subsection*{Acknowledgments}

We are grateful to T.~Hansmann, T.~Mannel, and T.~Ohl for valuable
discussions.  Particular thanks go to T.~Plehn and D.~Rainwater for a
critical reading of the manuscript.

\appendix
\section{The effective Lagrangian}

In this appendix we collect the dimension-six operators which are
present in the low-energy theory after integrating out the heavy
degrees of freedom.  The result can be written in the form
\begin{equation}\label{L6}
\begin{split}
  \LL_6 &= f_{JJ}^{(3)}\Op_{JJ}^{(3)} + f_{JJ}^{(1)}\Op_{JJ}^{(1)}
\\ &\quad
  + f_{VN}^{(1)} \Op_{VN}
  - \frac12(f_{VV}^{(3)}+2f_{VV}^{(1)})\Op_{J_S}'
\\ &\quad
  - 2(f_{VW}+2f_{VB})\Op_{BW}
\\ &\quad
  - 4f_{VW}\Op_W - 2f_{VW}\Op_{WW}'
\\ &\quad
  - 8f_{VB}\Op_B 
  - 4f_{VB}\Op_{BB}' 
\\ &\quad
  + f_{h,1}\Op_{h,1}' 
  + f_{hh}\Op_{hh}'
    + f_{h,3}\Op_{h,3}' 
\\ &\quad
  + f_{Vq} \Op_{Vq} + f_{Vt} \Op_{Vt} + f_{hq} \Op_{hq}'.
\end{split}
\end{equation}
For the operators, we adopt the basis of Ref.~\cite{Zep} with some minor
modifications.  The operators are defined as follows:
\begin{subequations}
  \begin{align}
    \Op^{(3)}_{JJ} &= \tr{J^{(3),\mu} J^{(3)}_\mu}, \\
    \Op_{JJ}^{(1)} &= J^{(1),\mu} J^{(1)}_\mu, \\
    \Op_{VN}^{(1)} &= V^{(1),\mu} J^{(1)}_{N,\mu}, \\
    \Op_{J_S}' &= (h^\dagger h-v^2/2)(h^\dagger J_S + J_S^\dagger h), \\
    \Op_{BW} &= -\frac12 B_{\mu\nu} h^\dagger W^{\mu\nu} h, \\
    \Op_W &= i(D_\mu h)^\dagger W^{\mu\nu} (D_\nu h), \\
    \Op_{WW}' &= -\frac12 (h^\dagger h - v^2/2)\tr{W_{\mu\nu} W^{\mu\nu}}, \\
    \Op_B  &= \frac{i}{2} (D_\mu h)^\dagger (D_\nu h) B^{\mu\nu}, \\
    \Op_{BB}' &= -\frac14 (h^\dagger h - v^2/2) B_{\mu\nu} B^{\mu\nu}, \\
    \Op_{h,1}' &= \left((D_\mu h)^\dagger h\right)
                  \left(h^\dagger(D^\mu h)\right)
                 - (v^2/2) (D_\mu h)^\dagger(D^\mu h), \\
    \Op_{hh}' &= (h^\dagger h-v^2/2)\left((D_\mu h)^\dagger (D^\mu h)\right), \\
    \Op_{h,3}' &= \frac13(h^\dagger h-v^2/2)^3, \\
    \Op_{Vq} &= \bar{\tilde q}_L\fmslash[-4mu]{V}^T \tilde q_L, \\
    \Op_{Vt} &= \bar{t}_R\fmslash[-4mu]{V}^{(1)} t_R, \\
    \Op_{hq}' &= (h^\dagger h- v^2/2)
                 \left(\bar t_R h^T {\tilde q}_L + \hc\right),
  \end{align}
\end{subequations}
where $V_\mu$ is the vector field
\begin{equation}
  V_\mu = i\left[h (D_\mu h)^\dagger - (D_\mu h) h^\dagger\right]
\end{equation}
with triplet and singlet parts
\begin{equation}
  V_\mu^{(1)} = \tr{V_\mu},
\quad
  V_\mu^{(3)} = V_\mu - \frac12\tr{V_\mu}.
\end{equation}
The triplet fermion current $J^{(3)}$ is the usual isospin current,
$J^{(1)}_Y$ is the hypercharge current, and $J_S$ is the fermion
current coupled to the SM Higgs doublet.  The exact form of the
currents $J^{(1)}$ and $J^{(1)}_N$ is model-dependent.  In Little
Higgs models where only one $U(1)$ gauge boson is coupled to fermions,
$J_N$ vanishes, and $J^{(1)}$ is proportional to the hypercharge current.

In the Littlest Higgs model, the values of the coefficients
in~(\ref{L6}) are
\begin{subequations}
  \begin{align}
    f^{(3)}_{JJ} &= -\frac{4c^4}{F^2}, \\
    f_{JJ}^{(1)} &= -\frac{10}{F^2}, \\
    f_{VN}^{(1)} &= \frac{5(c^\pp-s^\pp)}{F^2}, \\
    f^{(3)}_{VV} &= \frac{(1+2c^2)(c^2-s^2)}{4F^2} - \frac{1}{6F^2}, \\
    f_{VV}^{(1)} &= \frac{5(1+2c^\pp-4a)(c^\pp-s^\pp)}{8F^2}, \\
    f_{VW} &=  -\frac{1}{2g^2}f_{VJ}^{(3)}
           = -\frac{c^2(c^2-s^2)}{g^2F^2}, \\
    f_{VB} &=  -\frac{1}{2g^\pp}f_{VJ}^{(1)}
           = -\frac{5(c^\pp-a)(c^\pp-s^\pp)}{2g^\pp F^2}, \\
    f_{h,1} &= 4f_{VV}^{(1)} + \frac{4\lambda_{2\phi}^2}{M_\phi^4}, \\
    f_{hh} &= 3f_{VV}^{(3)}+2f_{VV}^{(1)} 
          + \frac{4\lambda_{2\phi}^2}{M_\phi^4}, \\
    f_{h,3} &= - 3\frac{m_H^2}{v^2}
            \left(f_{VV}^{(3)} + 2f_{VV}^{(1)} - \frac{1}{3F^2}\right),
  \end{align}
  \begin{align}
    f_{Vq} &= -\frac{s_t^4}{F^2}, \\
    f_{Vt} &= 0, \\
    f_{hq} &= \frac{\sqrt2\lambda_t}{F^2}
              \left(c_t^2 s_t^2 - \frac{2}{3} 
                    + \frac{\sqrt2\lambda_{2\phi}F}{M_\phi^2}\right),
  \end{align}
\end{subequations}
while the expressions for the Coleman-Weinberg potential parameters
are given in (\ref{Mphi2}--\ref{pot-l6}).  The singlet current
$J^{(1)}$ is given by
\begin{equation}
  J^{(1)}_{\mu} = (c^\pp - a)J^{(1)}_{Y,\mu} + J^{(1)}_{N,\mu},
\end{equation}
where $J_{N}$ collects the terms that cannot be absorbed into the
hypercharge current by shifting the parameter~$a$.  In the original
version of the model~\cite{LHmin}, $J_N$ and $a$ both vanish.


\end{document}